\documentclass[submission,copyright,creativecommons]{eptcs}
\providecommand{\event}{MSFP 2022} % Name of the event you are submitting to

\usepackage[utf8x]{inputenc}
\usepackage{amsmath}
\usepackage{amsfonts}
\usepackage{amssymb}
\usepackage[svgnames]{xcolor} % Provides better colors with upper cases
\usepackage{agda} % The Agda package to input Agda code
\usepackage{lineno} % Allows to number lines
\usepackage{xargs} % Allows to create better commands
\usepackage{tikz}
\usepackage{xspace}
\usetikzlibrary{positioning,backgrounds,arrows,automata}
\usepackage{multicol}
\usepackage{mdframed}
\usepackage{fancyvrb}
\usepackage{listings}

\newcommand{\coq}{\textsc{Coq}\xspace}
\newcommand{\agda}{\textsc{Agda}\xspace}
\newcommand{\lagda}{\texttt{lagda}\xspace}
\newcommand{\haskell}{\textsc{Haskell}\xspace}
\newcommand{\caml}{\textsc{OCaml}\xspace}
\newcommand{\agsy}{\textsc{Agsy}\xspace}
\newcommand{\datatypes}{datatypes\xspace}

\newcommand{\datatype}{datatype\xspace}
\newcommand{\libName}{\textsc{LibNDT}\xspace}
\newcommand{\linear}{\textsc{LNDT}\xspace}
\newcommand{\linears}{\textsc{LNDTs}\xspace}
\newcommand{\linearsFull}{linked nested datatypes\xspace}
\newcommand{\ttrans}{type transformer\xspace}
\newcommand{\ttranss}{type transformers\xspace}

\newcommand{\gitURL}{\url{https://github.com/mmontin/libndt}}

\DeclareUnicodeCharacter{8759}{\ensuremath{::}}
\DeclareUnicodeCharacter{955}{\ensuremath{\lambda}}
\DeclareUnicodeCharacter{969}{\ensuremath{\omega}}

\AgdaNoSpaceAroundCode % The name says it all
\renewcommandx*{\hrulefill}[2][1=0.3mm,2=0pt] % A \rule with parameterized positioning
    {\leavevmode \leaders \hbox to 1pt{\rule[#2]{1pt}{#1}} \hfill \kern 0pt}

\tikzset{
    node distance=5mm and -9mm, %Caractéristiques sur les noeuds
    boite/.style={ %Caractéristiques sur le styles des boites
        fill=blue!50!cyan!70,
        draw,
        align=center,
        font=\footnotesize\bfseries,
        text=white,
    },
    boite coins ronds/.style={ %Caractéristiques sur les boites
        boite,
        rounded corners=3pt
    },
    boite circulaire/.style={ %Caractéristiques sur les boites circulaires
        boite,
        circle
    },
    fleche/.style={ %Caractéristiques sur les flèches
        line cap=round,
        -latex,
        line width=0.25mm,
	    draw=blue!50!cyan!30,
	},
	grosse fleche/.style={ %Caractéristiques sur les grosses flèches
	    fleche,
	    line width=1mm
	},
}

\definecolor{beige}{rgb}{1.0, 0.99, 0.82}
\definecolor{orange_coq}{RGB}{255,69,0}
\definecolor{green_coq}{RGB}{0,100,0}
\definecolor{blue_type}{RGB}{0,128,128}
\definecolor{brown_coq}{RGB}{88,41,0}

\lstdefinelanguage{Coq}
{
	extendedchars=true,
	numbers=none,
	numberstyle={},
	tabsize=2,
	basicstyle=\small,
	sensitive=true,
	backgroundcolor=\color{beige},
	keywords={[0]Theorem, Lemma, Record, Inductive, Record, Fixpoint, Function, Definition, Axiom, Variable, Hypothesis},
	keywordstyle={[0]\color{orange_coq} \bfseries},  
	keywords={[2]forall, if, then, else, match, with, end, let, in },
	keywordstyle={[2]\color{green_coq}},
	keywords={[4]Type, Prop, Set},
	keywordstyle={[4]\color{blue_type}\bfseries},
	keywords={[5]Proof, Qed, Defined, Admitted, Abort, Print, Check, Eval, Require, Import,
		+, -, *, {, }}, 
	keywordstyle={[5]\color{violet} \bfseries},
	morecomment=[l]{(*},
	commentstyle={\itshape\color{brown_coq}\bfseries},
	string=[b]{"},
	stringstyle=\color{red},
	showstringspaces=false,
	literate={ -> }{$\rightarrow$}1 { <- }{$\leftarrow$}1 
	{<-> }{$\leftrightarrow$}1
	{\\/}{$\lor$}1 {/\\}{$\land$}1
	{~}{$\sim$}1,
	aboveskip=4mm,
	belowskip=4mm,
	breakatwhitespace=false,
	breaklines=true,
	captionpos=b,
	escapeinside={\%*}{*)},
	extendedchars=true,  
	framexleftmargin=1pt,
	framextopmargin=3pt,
	framexbottommargin=6pt,
	frame=tb,
	keepspaces=true
}

\newcommand{\exampleList}{
\begin{code}%
\>[0]\AgdaKeyword{data}\AgdaSpace{}%
\AgdaDatatype{List₀}\AgdaSpace{}%
\AgdaSymbol{\{}\AgdaBound{a}\AgdaSymbol{\}}\AgdaSpace{}%
\AgdaSymbol{(}\AgdaBound{A}\AgdaSpace{}%
\AgdaSymbol{:}\AgdaSpace{}%
\AgdaPrimitive{Set}\AgdaSpace{}%
\AgdaBound{a}\AgdaSymbol{)}\AgdaSpace{}%
\AgdaSymbol{:}\AgdaSpace{}%
\AgdaPrimitive{Set}\AgdaSpace{}%
\AgdaBound{a}\AgdaSpace{}%
\AgdaKeyword{where}\<%
\\
\>[0][@{}l@{\AgdaIndent{0}}]%
\>[2]\AgdaInductiveConstructor{[]}\AgdaSpace{}%
\AgdaSymbol{:}\AgdaSpace{}%
\AgdaDatatype{List₀}\AgdaSpace{}%
\AgdaBound{A}\<%
\\
\>[2]\AgdaOperator{\AgdaInductiveConstructor{\AgdaUnderscore{}∷\AgdaUnderscore{}}}\AgdaSpace{}%
\AgdaSymbol{:}\AgdaSpace{}%
\AgdaBound{A}\AgdaSpace{}%
\AgdaSymbol{→}\AgdaSpace{}%
\AgdaDatatype{List₀}\AgdaSpace{}%
\AgdaBound{A}\AgdaSpace{}%
\AgdaSymbol{→}\AgdaSpace{}%
\AgdaDatatype{List₀}\AgdaSpace{}%
\AgdaBound{A}\<%
\end{code}
}

\newcommand{\exampleNest}{
\begin{code}%
\>[0]\AgdaKeyword{data}\AgdaSpace{}%
\AgdaDatatype{Nest₀}\AgdaSpace{}%
\AgdaSymbol{\{}\AgdaBound{a}\AgdaSymbol{\}}\AgdaSpace{}%
\AgdaSymbol{(}\AgdaBound{A}\AgdaSpace{}%
\AgdaSymbol{:}\AgdaSpace{}%
\AgdaPrimitive{Set}\AgdaSpace{}%
\AgdaBound{a}\AgdaSymbol{)}\AgdaSpace{}%
\AgdaSymbol{:}\AgdaSpace{}%
\AgdaPrimitive{Set}\AgdaSpace{}%
\AgdaBound{a}\AgdaSpace{}%
\AgdaKeyword{where}\<%
\\
\>[0][@{}l@{\AgdaIndent{0}}]%
\>[2]\AgdaInductiveConstructor{[]}\AgdaSpace{}%
\AgdaSymbol{:}\AgdaSpace{}%
\AgdaDatatype{Nest₀}\AgdaSpace{}%
\AgdaBound{A}\<%
\\
\>[2]\AgdaOperator{\AgdaInductiveConstructor{\AgdaUnderscore{}∷\AgdaUnderscore{}}}\AgdaSpace{}%
\AgdaSymbol{:}\AgdaSpace{}%
\AgdaBound{A}\AgdaSpace{}%
\AgdaSymbol{→}\AgdaSpace{}%
\AgdaDatatype{Nest₀}\AgdaSpace{}%
\AgdaSymbol{(}\AgdaBound{A}\AgdaSpace{}%
\AgdaOperator{\AgdaFunction{×}}\AgdaSpace{}%
\AgdaBound{A}\AgdaSymbol{)}\AgdaSpace{}%
\AgdaSymbol{→}\AgdaSpace{}%
\AgdaDatatype{Nest₀}\AgdaSpace{}%
\AgdaBound{A}\<%
\end{code}
}

\newcommand{\exampleBush}{
\begin{code}%
\>[0]\AgdaKeyword{data}\AgdaSpace{}%
\AgdaDatatype{Bush₀}\AgdaSpace{}%
\AgdaSymbol{\{}\AgdaBound{a}\AgdaSymbol{\}}\AgdaSpace{}%
\AgdaSymbol{(}\AgdaBound{A}\AgdaSpace{}%
\AgdaSymbol{:}\AgdaSpace{}%
\AgdaPrimitive{Set}\AgdaSpace{}%
\AgdaBound{a}\AgdaSymbol{)}\AgdaSpace{}%
\AgdaSymbol{:}\AgdaSpace{}%
\AgdaPrimitive{Set}\AgdaSpace{}%
\AgdaBound{a}\AgdaSpace{}%
\AgdaKeyword{where}\<%
\\
\>[0][@{}l@{\AgdaIndent{0}}]%
\>[2]\AgdaInductiveConstructor{[]}\AgdaSpace{}%
\AgdaSymbol{:}\AgdaSpace{}%
\AgdaDatatype{Bush₀}\AgdaSpace{}%
\AgdaBound{A}\<%
\\
\>[2]\AgdaOperator{\AgdaInductiveConstructor{\AgdaUnderscore{}∷\AgdaUnderscore{}}}\AgdaSpace{}%
\AgdaSymbol{:}\AgdaSpace{}%
\AgdaBound{A}\AgdaSpace{}%
\AgdaSymbol{→}\AgdaSpace{}%
\AgdaDatatype{Bush₀}\AgdaSpace{}%
\AgdaSymbol{(}\AgdaDatatype{Bush₀}\AgdaSpace{}%
\AgdaBound{A}\AgdaSymbol{)}\AgdaSpace{}%
\AgdaSymbol{→}\AgdaSpace{}%
\AgdaDatatype{Bush₀}\AgdaSpace{}%
\AgdaBound{A}\<%
\end{code}
}

\newcommand{\TTdef}{
\begin{code}%
\>[0]\AgdaFunction{TT}\AgdaSpace{}%
\AgdaSymbol{:}\AgdaSpace{}%
\AgdaPrimitive{Setω}\<%
\\
\>[0]\AgdaFunction{TT}\AgdaSpace{}%
\AgdaSymbol{=}\AgdaSpace{}%
\AgdaSymbol{∀}\AgdaSpace{}%
\AgdaSymbol{\{}\AgdaBound{a}\AgdaSymbol{\}}\AgdaSpace{}%
\AgdaSymbol{→}\AgdaSpace{}%
\AgdaPrimitive{Set}\AgdaSpace{}%
\AgdaBound{a}\AgdaSpace{}%
\AgdaSymbol{→}\AgdaSpace{}%
\AgdaPrimitive{Set}\AgdaSpace{}%
\AgdaBound{a}\<%
\end{code}
}

\newcommand{\lndt}{
\begin{code}%
\>[0]\AgdaKeyword{data}\AgdaSpace{}%
\AgdaDatatype{LNDT}\AgdaSpace{}%
\AgdaSymbol{(}\AgdaBound{F}\AgdaSpace{}%
\AgdaSymbol{:}\AgdaSpace{}%
\AgdaFunction{TT}\AgdaSymbol{)}\AgdaSpace{}%
\AgdaSymbol{\{}\AgdaBound{a}\AgdaSymbol{\}}\AgdaSpace{}%
\AgdaSymbol{(}\AgdaBound{A}\AgdaSpace{}%
\AgdaSymbol{:}\AgdaSpace{}%
\AgdaPrimitive{Set}\AgdaSpace{}%
\AgdaBound{a}\AgdaSymbol{)}\AgdaSpace{}%
\AgdaSymbol{:}\AgdaSpace{}%
\AgdaPrimitive{Set}\AgdaSpace{}%
\AgdaBound{a}\AgdaSpace{}%
\AgdaKeyword{where}\<%
\\
\>[0][@{}l@{\AgdaIndent{0}}]%
\>[2]\AgdaInductiveConstructor{[]}\AgdaSpace{}%
\AgdaSymbol{:}\AgdaSpace{}%
\AgdaDatatype{LNDT}\AgdaSpace{}%
\AgdaBound{F}\AgdaSpace{}%
\AgdaBound{A}\<%
\\
\>[2]\AgdaOperator{\AgdaInductiveConstructor{\AgdaUnderscore{}∷\AgdaUnderscore{}}}\AgdaSpace{}%
\AgdaSymbol{:}\AgdaSpace{}%
\AgdaBound{A}\AgdaSpace{}%
\AgdaSymbol{→}\AgdaSpace{}%
\AgdaDatatype{LNDT}\AgdaSpace{}%
\AgdaBound{F}\AgdaSpace{}%
\AgdaSymbol{(}\AgdaBound{F}\AgdaSpace{}%
\AgdaBound{A}\AgdaSymbol{)}\AgdaSpace{}%
\AgdaSymbol{→}\AgdaSpace{}%
\AgdaDatatype{LNDT}\AgdaSpace{}%
\AgdaBound{F}\AgdaSpace{}%
\AgdaBound{A}\<%
\end{code}
}

\newcommand{\tuples}{
\begin{code}%
\>[0]\AgdaFunction{Tuple}\AgdaSpace{}%
\AgdaSymbol{:}\AgdaSpace{}%
\AgdaDatatype{ℕ}\AgdaSpace{}%
\AgdaSymbol{→}\AgdaSpace{}%
\AgdaFunction{TT}\<%
\\
\>[0]\AgdaFunction{Tuple}\AgdaSpace{}%
\AgdaInductiveConstructor{zero}\AgdaSpace{}%
\AgdaSymbol{=}\AgdaSpace{}%
\AgdaFunction{id}\AgdaSpace{}%
\AgdaComment{--\ id\ is\ the\ identity\ function}\<%
\\
\>[0]\AgdaFunction{Tuple}\AgdaSpace{}%
\AgdaSymbol{(}\AgdaInductiveConstructor{suc}\AgdaSpace{}%
\AgdaBound{n}\AgdaSymbol{)}\AgdaSpace{}%
\AgdaBound{A}\AgdaSpace{}%
\AgdaSymbol{=}\AgdaSpace{}%
\AgdaBound{A}\AgdaSpace{}%
\AgdaOperator{\AgdaFunction{×}}\AgdaSpace{}%
\AgdaSymbol{(}\AgdaFunction{Tuple}\AgdaSpace{}%
\AgdaBound{n}\AgdaSpace{}%
\AgdaBound{A}\AgdaSymbol{)}\<%
\end{code}
}

\newcommand{\nary}{
\begin{code}%
\>[0]\AgdaFunction{N-PT}\AgdaSpace{}%
\AgdaSymbol{:}\AgdaSpace{}%
\AgdaDatatype{ℕ}\AgdaSpace{}%
\AgdaSymbol{→}\AgdaSpace{}%
\AgdaFunction{TT}\<%
\\
\>[0]\AgdaFunction{N-PT}\AgdaSpace{}%
\AgdaBound{n}\AgdaSpace{}%
\AgdaSymbol{=}\AgdaSpace{}%
\AgdaDatatype{LNDT}\AgdaSpace{}%
\AgdaSymbol{(}\AgdaFunction{Tuple}\AgdaSpace{}%
\AgdaBound{n}\AgdaSymbol{)}\<%
\end{code}
}

\newcommand{\lndtdefs}{
\begin{multicols}{3}
\begin{code}%
\>[0]\AgdaFunction{List}\AgdaSpace{}%
\AgdaSymbol{:}\AgdaSpace{}%
\AgdaFunction{TT}\<%
\\
\>[0]\AgdaFunction{List}\AgdaSpace{}%
\AgdaSymbol{=}\AgdaSpace{}%
\AgdaFunction{N-PT}\AgdaSpace{}%
\AgdaNumber{0}\<%
\end{code}
\columnbreak
\begin{code}%
\>[0]\AgdaFunction{Nest}\AgdaSpace{}%
\AgdaSymbol{:}\AgdaSpace{}%
\AgdaFunction{TT}\<%
\\
\>[0]\AgdaFunction{Nest}\AgdaSpace{}%
\AgdaSymbol{=}\AgdaSpace{}%
\AgdaFunction{N-PT}\AgdaSpace{}%
\AgdaNumber{1}\<%
\end{code}
\columnbreak
\begin{code}[hide]%
\>[0]\AgdaSymbol{\{-\#}\AgdaSpace{}%
\AgdaKeyword{TERMINATING}\AgdaSpace{}%
\AgdaSymbol{\#-\}}\<%
\end{code}
\begin{code}%
\>[0]\AgdaFunction{Bush}\AgdaSpace{}%
\AgdaSymbol{:}\AgdaSpace{}%
\AgdaFunction{TT}\<%
\\
\>[0]\AgdaFunction{Bush}\AgdaSpace{}%
\AgdaSymbol{=}\AgdaSpace{}%
\AgdaDatatype{LNDT}\AgdaSpace{}%
\AgdaFunction{Bush}\<%
\end{code}
\end{multicols}
}

\newcommand{\examplelist}{
\begin{code}%
\>[0]\AgdaFunction{list-example}\AgdaSpace{}%
\AgdaSymbol{:}\AgdaSpace{}%
\AgdaFunction{List}\AgdaSpace{}%
\AgdaDatatype{ℕ}\<%
\\
\>[0]\AgdaFunction{list-example}\AgdaSpace{}%
\AgdaSymbol{=}\AgdaSpace{}%
\AgdaNumber{2}\AgdaSpace{}%
\AgdaOperator{\AgdaInductiveConstructor{∷}}\AgdaSpace{}%
\AgdaNumber{3}\AgdaSpace{}%
\AgdaOperator{\AgdaInductiveConstructor{∷}}\AgdaSpace{}%
\AgdaNumber{7}\AgdaSpace{}%
\AgdaOperator{\AgdaInductiveConstructor{∷}}\AgdaSpace{}%
\AgdaNumber{11}\AgdaSpace{}%
\AgdaOperator{\AgdaInductiveConstructor{∷}}\AgdaSpace{}%
\AgdaNumber{13}\AgdaSpace{}%
\AgdaOperator{\AgdaInductiveConstructor{∷}}\AgdaSpace{}%
\AgdaNumber{17}\AgdaSpace{}%
\AgdaOperator{\AgdaInductiveConstructor{∷}}\AgdaSpace{}%
\AgdaNumber{19}\AgdaSpace{}%
\AgdaOperator{\AgdaInductiveConstructor{∷}}\AgdaSpace{}%
\AgdaNumber{23}\AgdaSpace{}%
\AgdaOperator{\AgdaInductiveConstructor{∷}}\AgdaSpace{}%
\AgdaNumber{29}\AgdaSpace{}%
\AgdaOperator{\AgdaInductiveConstructor{∷}}\AgdaSpace{}%
\AgdaNumber{31}\AgdaSpace{}%
\AgdaOperator{\AgdaInductiveConstructor{∷}}\AgdaSpace{}%
\AgdaNumber{37}\AgdaSpace{}%
\AgdaOperator{\AgdaInductiveConstructor{∷}}\AgdaSpace{}%
\AgdaNumber{41}\AgdaSpace{}%
\AgdaOperator{\AgdaInductiveConstructor{∷}}\AgdaSpace{}%
\AgdaNumber{43}\AgdaSpace{}%
\AgdaOperator{\AgdaInductiveConstructor{∷}}\AgdaSpace{}%
\AgdaNumber{47}\AgdaSpace{}%
\AgdaOperator{\AgdaInductiveConstructor{∷}}\AgdaSpace{}%
\AgdaNumber{53}\AgdaSpace{}%
\AgdaOperator{\AgdaInductiveConstructor{∷}}\AgdaSpace{}%
\AgdaNumber{59}\AgdaSpace{}%
\AgdaOperator{\AgdaInductiveConstructor{∷}}\AgdaSpace{}%
\AgdaInductiveConstructor{[]}\<%
\end{code}
}

\newcommand{\examplebush}{
\begin{code}%
\>[0]\AgdaFunction{bush-example}\AgdaSpace{}%
\AgdaSymbol{:}\AgdaSpace{}%
\AgdaFunction{Bush}\AgdaSpace{}%
\AgdaDatatype{ℕ}\<%
\\
\>[0]\AgdaFunction{bush-example}\AgdaSpace{}%
\AgdaSymbol{=}\AgdaSpace{}%
\AgdaNumber{5}\AgdaSpace{}%
\AgdaOperator{\AgdaInductiveConstructor{∷}}\AgdaSpace{}%
\AgdaSymbol{(}\AgdaNumber{16}\AgdaSpace{}%
\AgdaOperator{\AgdaInductiveConstructor{∷}}\AgdaSpace{}%
\AgdaInductiveConstructor{[]}\AgdaSymbol{)}\AgdaSpace{}%
\AgdaOperator{\AgdaInductiveConstructor{∷}}\AgdaSpace{}%
\AgdaSymbol{((}\AgdaNumber{87}\AgdaSpace{}%
\AgdaOperator{\AgdaInductiveConstructor{∷}}\AgdaSpace{}%
\AgdaInductiveConstructor{[]}\AgdaSymbol{)}\AgdaSpace{}%
\AgdaOperator{\AgdaInductiveConstructor{∷}}\AgdaSpace{}%
\AgdaSymbol{((}\AgdaNumber{56}\AgdaSpace{}%
\AgdaOperator{\AgdaInductiveConstructor{∷}}\AgdaSpace{}%
\AgdaInductiveConstructor{[]}\AgdaSymbol{)}\AgdaSpace{}%
\AgdaOperator{\AgdaInductiveConstructor{∷}}\AgdaSpace{}%
\AgdaInductiveConstructor{[]}\AgdaSymbol{)}\AgdaSpace{}%
\AgdaOperator{\AgdaInductiveConstructor{∷}}\AgdaSpace{}%
\AgdaInductiveConstructor{[]}\AgdaSymbol{)}\AgdaSpace{}%
\AgdaOperator{\AgdaInductiveConstructor{∷}}\AgdaSpace{}%
\AgdaInductiveConstructor{[]}\<%
\end{code}
}
\newcommand{\examplenest}{
\begin{code}%
\>[0]\AgdaFunction{nest-example}\AgdaSpace{}%
\AgdaSymbol{:}\AgdaSpace{}%
\AgdaFunction{Nest}\AgdaSpace{}%
\AgdaDatatype{ℕ}\<%
\\
\>[0]\AgdaFunction{nest-example}\AgdaSpace{}%
\AgdaSymbol{=}\AgdaSpace{}%
\AgdaNumber{4}\AgdaSpace{}%
\AgdaOperator{\AgdaInductiveConstructor{∷}}\AgdaSpace{}%
\AgdaSymbol{(}\AgdaNumber{4}\AgdaSpace{}%
\AgdaOperator{\AgdaInductiveConstructor{,}}\AgdaSpace{}%
\AgdaNumber{5}\AgdaSymbol{)}\AgdaSpace{}%
\AgdaOperator{\AgdaInductiveConstructor{∷}}\AgdaSpace{}%
\AgdaSymbol{((}\AgdaNumber{6}\AgdaSpace{}%
\AgdaOperator{\AgdaInductiveConstructor{,}}\AgdaSpace{}%
\AgdaNumber{56}\AgdaSymbol{)}\AgdaSpace{}%
\AgdaOperator{\AgdaInductiveConstructor{,}}\AgdaSpace{}%
\AgdaSymbol{(}\AgdaNumber{81}\AgdaSpace{}%
\AgdaOperator{\AgdaInductiveConstructor{,}}\AgdaSpace{}%
\AgdaNumber{7}\AgdaSymbol{))}\AgdaSpace{}%
\AgdaOperator{\AgdaInductiveConstructor{∷}}\AgdaSpace{}%
\AgdaSymbol{(((}\AgdaNumber{5}\AgdaSpace{}%
\AgdaOperator{\AgdaInductiveConstructor{,}}\AgdaSpace{}%
\AgdaNumber{4}\AgdaSymbol{)}\AgdaSpace{}%
\AgdaOperator{\AgdaInductiveConstructor{,}}\AgdaSpace{}%
\AgdaSymbol{(}\AgdaNumber{1}\AgdaSpace{}%
\AgdaOperator{\AgdaInductiveConstructor{,}}\AgdaSpace{}%
\AgdaNumber{7}\AgdaSymbol{))}\AgdaSpace{}%
\AgdaOperator{\AgdaInductiveConstructor{,}}\AgdaSpace{}%
\AgdaSymbol{((}\AgdaNumber{7}\AgdaSpace{}%
\AgdaOperator{\AgdaInductiveConstructor{,}}\AgdaSpace{}%
\AgdaNumber{6}\AgdaSymbol{)}\AgdaSpace{}%
\AgdaOperator{\AgdaInductiveConstructor{,}}\AgdaSpace{}%
\AgdaNumber{4}\AgdaSpace{}%
\AgdaOperator{\AgdaInductiveConstructor{,}}\AgdaSpace{}%
\AgdaNumber{7}\AgdaSymbol{))}\AgdaSpace{}%
\AgdaOperator{\AgdaInductiveConstructor{∷}}\AgdaSpace{}%
\AgdaInductiveConstructor{[]}\<%
\end{code}
}

\newcommand{\lndtnull}{
\begin{multicols}{2}
\begin{code}%
\>[0]\AgdaKeyword{data}\AgdaSpace{}%
\AgdaDatatype{⊥}\AgdaSpace{}%
\AgdaSymbol{\{}\AgdaBound{a}\AgdaSymbol{\}}\AgdaSpace{}%
\AgdaSymbol{:}\AgdaSpace{}%
\AgdaPrimitive{Set}\AgdaSpace{}%
\AgdaBound{a}\AgdaSpace{}%
\AgdaKeyword{where}\<%
\\
\>[0][@{}l@{\AgdaIndent{0}}]%
\>[2]\AgdaComment{--\ Empty\ type}\<%
\end{code}
\columnbreak
\begin{code}%
\>[0]\AgdaFunction{Null}\AgdaSpace{}%
\AgdaSymbol{:}\AgdaSpace{}%
\AgdaFunction{TT}\<%
\\
\>[0]\AgdaFunction{Null}\AgdaSpace{}%
\AgdaSymbol{\AgdaUnderscore{}}\AgdaSpace{}%
\AgdaSymbol{=}\AgdaSpace{}%
\AgdaDatatype{⊥}\<%
\end{code}
\end{multicols}
}

\newcommand{\maybe}{
\begin{multicols}{2}
\begin{code}%
\>[0]\AgdaFunction{Maybe}\AgdaSpace{}%
\AgdaSymbol{:}\AgdaSpace{}%
\AgdaFunction{TT}\<%
\\
\>[0]\AgdaFunction{Maybe}\AgdaSpace{}%
\AgdaSymbol{=}\AgdaSpace{}%
\AgdaDatatype{LNDT}\AgdaSpace{}%
\AgdaFunction{Null}\<%
\end{code}
\columnbreak
\begin{code}%
\>[0]\AgdaKeyword{pattern}\AgdaSpace{}%
\AgdaInductiveConstructor{nothing}\AgdaSpace{}%
\AgdaSymbol{=}\AgdaSpace{}%
\AgdaInductiveConstructor{[]}\<%
\\
\>[0]\AgdaKeyword{pattern}\AgdaSpace{}%
\AgdaInductiveConstructor{just}\AgdaSpace{}%
\AgdaBound{x}%
\>[16]\AgdaSymbol{=}\AgdaSpace{}%
\AgdaBound{x}\AgdaSpace{}%
\AgdaOperator{\AgdaInductiveConstructor{∷}}\AgdaSpace{}%
\AgdaInductiveConstructor{[]}\<%
\end{code}
\end{multicols}
}

\newcommand{\listmap}{
\begin{code}%
\>[0]\AgdaFunction{list-map₀}\AgdaSpace{}%
\AgdaSymbol{:}\AgdaSpace{}%
\AgdaSymbol{∀}\AgdaSpace{}%
\AgdaSymbol{\{}\AgdaBound{a}\AgdaSpace{}%
\AgdaBound{b}\AgdaSymbol{\}}\AgdaSpace{}%
\AgdaSymbol{\{}\AgdaBound{A}\AgdaSpace{}%
\AgdaSymbol{:}\AgdaSpace{}%
\AgdaPrimitive{Set}\AgdaSpace{}%
\AgdaBound{a}\AgdaSymbol{\}}\AgdaSpace{}%
\AgdaSymbol{\{}\AgdaBound{B}\AgdaSpace{}%
\AgdaSymbol{:}\AgdaSpace{}%
\AgdaPrimitive{Set}\AgdaSpace{}%
\AgdaBound{b}\AgdaSymbol{\}}\AgdaSpace{}%
\AgdaSymbol{→}\AgdaSpace{}%
\AgdaSymbol{(}\AgdaBound{A}\AgdaSpace{}%
\AgdaSymbol{→}\AgdaSpace{}%
\AgdaBound{B}\AgdaSymbol{)}\AgdaSpace{}%
\AgdaSymbol{→}\AgdaSpace{}%
\AgdaFunction{List}\AgdaSpace{}%
\AgdaBound{A}\AgdaSpace{}%
\AgdaSymbol{→}\AgdaSpace{}%
\AgdaFunction{List}\AgdaSpace{}%
\AgdaBound{B}\<%
\\
\>[0]\AgdaFunction{list-map₀}\AgdaSpace{}%
\AgdaBound{f}\AgdaSpace{}%
\AgdaInductiveConstructor{[]}\AgdaSpace{}%
\AgdaSymbol{=}\AgdaSpace{}%
\AgdaInductiveConstructor{[]}\<%
\\
\>[0]\AgdaFunction{list-map₀}\AgdaSpace{}%
\AgdaBound{f}\AgdaSpace{}%
\AgdaSymbol{(}\AgdaBound{x}\AgdaSpace{}%
\AgdaOperator{\AgdaInductiveConstructor{∷}}\AgdaSpace{}%
\AgdaBound{l}\AgdaSymbol{)}\AgdaSpace{}%
\AgdaSymbol{=}\AgdaSpace{}%
\AgdaBound{f}\AgdaSpace{}%
\AgdaBound{x}\AgdaSpace{}%
\AgdaOperator{\AgdaInductiveConstructor{∷}}\AgdaSpace{}%
\AgdaFunction{list-map₀}\AgdaSpace{}%
\AgdaBound{f}\AgdaSpace{}%
\AgdaBound{l}\<%
\end{code}
}

\newcommand{\nestmap}{
\begin{code}%
\>[0]\AgdaFunction{nest-map₀}\AgdaSpace{}%
\AgdaSymbol{:}\AgdaSpace{}%
\AgdaSymbol{∀}\AgdaSpace{}%
\AgdaSymbol{\{}\AgdaBound{a}\AgdaSpace{}%
\AgdaBound{b}\AgdaSymbol{\}}\AgdaSpace{}%
\AgdaSymbol{\{}\AgdaBound{A}\AgdaSpace{}%
\AgdaSymbol{:}\AgdaSpace{}%
\AgdaPrimitive{Set}\AgdaSpace{}%
\AgdaBound{a}\AgdaSymbol{\}}\AgdaSpace{}%
\AgdaSymbol{\{}\AgdaBound{B}\AgdaSpace{}%
\AgdaSymbol{:}\AgdaSpace{}%
\AgdaPrimitive{Set}\AgdaSpace{}%
\AgdaBound{b}\AgdaSymbol{\}}\AgdaSpace{}%
\AgdaSymbol{→}\AgdaSpace{}%
\AgdaSymbol{(}\AgdaBound{A}\AgdaSpace{}%
\AgdaSymbol{→}\AgdaSpace{}%
\AgdaBound{B}\AgdaSymbol{)}\AgdaSpace{}%
\AgdaSymbol{→}\AgdaSpace{}%
\AgdaFunction{Nest}\AgdaSpace{}%
\AgdaBound{A}\AgdaSpace{}%
\AgdaSymbol{→}\AgdaSpace{}%
\AgdaFunction{Nest}\AgdaSpace{}%
\AgdaBound{B}\<%
\\
\>[0]\AgdaFunction{nest-map₀}\AgdaSpace{}%
\AgdaBound{f}\AgdaSpace{}%
\AgdaInductiveConstructor{[]}\AgdaSpace{}%
\AgdaSymbol{=}\AgdaSpace{}%
\AgdaInductiveConstructor{[]}\<%
\\
\>[0]\AgdaFunction{nest-map₀}\AgdaSpace{}%
\AgdaBound{f}\AgdaSpace{}%
\AgdaSymbol{(}\AgdaBound{x}\AgdaSpace{}%
\AgdaOperator{\AgdaInductiveConstructor{∷}}\AgdaSpace{}%
\AgdaBound{n}\AgdaSymbol{)}\AgdaSpace{}%
\AgdaSymbol{=}\AgdaSpace{}%
\AgdaBound{f}\AgdaSpace{}%
\AgdaBound{x}\AgdaSpace{}%
\AgdaOperator{\AgdaInductiveConstructor{∷}}\AgdaSpace{}%
\AgdaFunction{nest-map₀}\AgdaSpace{}%
\AgdaSymbol{(λ}\AgdaSpace{}%
\AgdaSymbol{\{(}\AgdaBound{a}\AgdaSpace{}%
\AgdaOperator{\AgdaInductiveConstructor{,}}\AgdaSpace{}%
\AgdaBound{b}\AgdaSymbol{)}\AgdaSpace{}%
\AgdaSymbol{→}\AgdaSpace{}%
\AgdaSymbol{(}\AgdaBound{f}\AgdaSpace{}%
\AgdaBound{a}\AgdaSpace{}%
\AgdaOperator{\AgdaInductiveConstructor{,}}\AgdaSpace{}%
\AgdaBound{f}\AgdaSpace{}%
\AgdaBound{b}\AgdaSymbol{)\})}\AgdaSpace{}%
\AgdaBound{n}\<%
\end{code}
}

\newcommand{\lndtmapz}{
\begin{code}%
\>[0]\AgdaFunction{lndt-map₀}\AgdaSpace{}%
\AgdaSymbol{:}\AgdaSpace{}%
\AgdaSymbol{∀}\AgdaSpace{}%
\AgdaSymbol{\{}\AgdaBound{a}\AgdaSpace{}%
\AgdaBound{b}\AgdaSymbol{\}}\AgdaSpace{}%
\AgdaSymbol{\{}\AgdaBound{A}\AgdaSpace{}%
\AgdaSymbol{:}\AgdaSpace{}%
\AgdaPrimitive{Set}\AgdaSpace{}%
\AgdaBound{a}\AgdaSymbol{\}}\AgdaSpace{}%
\AgdaSymbol{\{}\AgdaBound{B}\AgdaSpace{}%
\AgdaSymbol{:}\AgdaSpace{}%
\AgdaPrimitive{Set}\AgdaSpace{}%
\AgdaBound{b}\AgdaSymbol{\}}\AgdaSpace{}%
\AgdaSymbol{\{}\AgdaBound{F}\AgdaSpace{}%
\AgdaSymbol{:}\AgdaSpace{}%
\AgdaFunction{TT}\AgdaSymbol{\}}\AgdaSpace{}%
\AgdaSymbol{→}\AgdaSpace{}%
\AgdaSymbol{(}\AgdaBound{A}\AgdaSpace{}%
\AgdaSymbol{→}\AgdaSpace{}%
\AgdaBound{B}\AgdaSymbol{)}\AgdaSpace{}%
\AgdaSymbol{→}\<%
\\
\>[0][@{}l@{\AgdaIndent{0}}]%
\>[2]\AgdaSymbol{(}\AgdaBound{T}\AgdaSpace{}%
\AgdaSymbol{:}\AgdaSpace{}%
\AgdaSymbol{∀}\AgdaSpace{}%
\AgdaSymbol{\{}\AgdaBound{a}\AgdaSpace{}%
\AgdaBound{b}\AgdaSymbol{\}}\AgdaSpace{}%
\AgdaSymbol{\{}\AgdaBound{A}\AgdaSpace{}%
\AgdaSymbol{:}\AgdaSpace{}%
\AgdaPrimitive{Set}\AgdaSpace{}%
\AgdaBound{a}\AgdaSymbol{\}}\AgdaSpace{}%
\AgdaSymbol{\{}\AgdaBound{B}\AgdaSpace{}%
\AgdaSymbol{:}\AgdaSpace{}%
\AgdaPrimitive{Set}\AgdaSpace{}%
\AgdaBound{b}\AgdaSymbol{\}}\AgdaSpace{}%
\AgdaSymbol{→}\AgdaSpace{}%
\AgdaSymbol{(}\AgdaBound{A}\AgdaSpace{}%
\AgdaSymbol{→}\AgdaSpace{}%
\AgdaBound{B}\AgdaSymbol{)}\AgdaSpace{}%
\AgdaSymbol{→}\AgdaSpace{}%
\AgdaSymbol{(}\AgdaBound{F}\AgdaSpace{}%
\AgdaBound{A}\AgdaSpace{}%
\AgdaSymbol{→}\AgdaSpace{}%
\AgdaBound{F}\AgdaSpace{}%
\AgdaBound{B}\AgdaSymbol{))}\<%
\\
\>[2]\AgdaSymbol{→}\AgdaSpace{}%
\AgdaDatatype{LNDT}\AgdaSpace{}%
\AgdaBound{F}\AgdaSpace{}%
\AgdaBound{A}\AgdaSpace{}%
\AgdaSymbol{→}\AgdaSpace{}%
\AgdaDatatype{LNDT}\AgdaSpace{}%
\AgdaBound{F}\AgdaSpace{}%
\AgdaBound{B}\<%
\\
\>[0]\AgdaFunction{lndt-map₀}\AgdaSpace{}%
\AgdaSymbol{\AgdaUnderscore{}}\AgdaSpace{}%
\AgdaSymbol{\AgdaUnderscore{}}\AgdaSpace{}%
\AgdaInductiveConstructor{[]}\AgdaSpace{}%
\AgdaSymbol{=}\AgdaSpace{}%
\AgdaInductiveConstructor{[]}\<%
\\
\>[0]\AgdaFunction{lndt-map₀}\AgdaSpace{}%
\AgdaBound{f}\AgdaSpace{}%
\AgdaBound{T}\AgdaSpace{}%
\AgdaSymbol{(}\AgdaBound{x}\AgdaSpace{}%
\AgdaOperator{\AgdaInductiveConstructor{∷}}\AgdaSpace{}%
\AgdaBound{e}\AgdaSymbol{)}\AgdaSpace{}%
\AgdaSymbol{=}\AgdaSpace{}%
\AgdaBound{f}\AgdaSpace{}%
\AgdaBound{x}\AgdaSpace{}%
\AgdaOperator{\AgdaInductiveConstructor{∷}}\AgdaSpace{}%
\AgdaFunction{lndt-map₀}\AgdaSpace{}%
\AgdaSymbol{(}\AgdaBound{T}\AgdaSpace{}%
\AgdaBound{f}\AgdaSymbol{)}\AgdaSpace{}%
\AgdaBound{T}\AgdaSpace{}%
\AgdaBound{e}\<%
\end{code}
}

\newcommand{\mappable}{
\begin{code}%
\>[0]\AgdaFunction{Map}\AgdaSpace{}%
\AgdaSymbol{:}\AgdaSpace{}%
\AgdaFunction{TT}\AgdaSpace{}%
\AgdaSymbol{→}\AgdaSpace{}%
\AgdaPrimitive{Setω}\<%
\\
\>[0]\AgdaFunction{Map}\AgdaSpace{}%
\AgdaBound{F}\AgdaSpace{}%
\AgdaSymbol{=}\AgdaSpace{}%
\AgdaSymbol{∀}\AgdaSpace{}%
\AgdaSymbol{\{}\AgdaBound{a}\AgdaSpace{}%
\AgdaBound{b}\AgdaSymbol{\}}\AgdaSpace{}%
\AgdaSymbol{\{}\AgdaBound{A}\AgdaSpace{}%
\AgdaSymbol{:}\AgdaSpace{}%
\AgdaPrimitive{Set}\AgdaSpace{}%
\AgdaBound{a}\AgdaSymbol{\}}\AgdaSpace{}%
\AgdaSymbol{\{}\AgdaBound{B}\AgdaSpace{}%
\AgdaSymbol{:}\AgdaSpace{}%
\AgdaPrimitive{Set}\AgdaSpace{}%
\AgdaBound{b}\AgdaSymbol{\}}\AgdaSpace{}%
\AgdaSymbol{→}\AgdaSpace{}%
\AgdaSymbol{(}\AgdaBound{A}\AgdaSpace{}%
\AgdaSymbol{→}\AgdaSpace{}%
\AgdaBound{B}\AgdaSymbol{)}\AgdaSpace{}%
\AgdaSymbol{→}\AgdaSpace{}%
\AgdaBound{F}\AgdaSpace{}%
\AgdaBound{A}\AgdaSpace{}%
\AgdaSymbol{→}\AgdaSpace{}%
\AgdaBound{F}\AgdaSpace{}%
\AgdaBound{B}\<%
\end{code}
}

\newcommand{\lndtmap}{
\begin{code}%
\>[0]\AgdaFunction{lndt-map}\AgdaSpace{}%
\AgdaSymbol{:}\AgdaSpace{}%
\AgdaSymbol{∀}\AgdaSpace{}%
\AgdaSymbol{\{}\AgdaBound{F}\AgdaSpace{}%
\AgdaSymbol{:}\AgdaSpace{}%
\AgdaFunction{TT}\AgdaSymbol{\}}\AgdaSpace{}%
\AgdaSymbol{→}\AgdaSpace{}%
\AgdaFunction{Map}\AgdaSpace{}%
\AgdaBound{F}\AgdaSpace{}%
\AgdaSymbol{→}\AgdaSpace{}%
\AgdaFunction{Map}\AgdaSpace{}%
\AgdaSymbol{(}\AgdaDatatype{LNDT}\AgdaSpace{}%
\AgdaBound{F}\AgdaSymbol{)}\<%
\\
\>[0]\AgdaFunction{lndt-map}\AgdaSpace{}%
\AgdaBound{F}\AgdaSpace{}%
\AgdaBound{f}\AgdaSpace{}%
\AgdaInductiveConstructor{[]}\AgdaSpace{}%
\AgdaSymbol{=}\AgdaSpace{}%
\AgdaInductiveConstructor{[]}\<%
\\
\>[0]\AgdaFunction{lndt-map}\AgdaSpace{}%
\AgdaBound{F}\AgdaSpace{}%
\AgdaBound{f}\AgdaSpace{}%
\AgdaSymbol{(}\AgdaBound{x}\AgdaSpace{}%
\AgdaOperator{\AgdaInductiveConstructor{∷}}\AgdaSpace{}%
\AgdaBound{e}\AgdaSymbol{)}\AgdaSpace{}%
\AgdaSymbol{=}\AgdaSpace{}%
\AgdaBound{f}\AgdaSpace{}%
\AgdaBound{x}\AgdaSpace{}%
\AgdaOperator{\AgdaInductiveConstructor{∷}}\AgdaSpace{}%
\AgdaFunction{lndt-map}\AgdaSpace{}%
\AgdaBound{F}\AgdaSpace{}%
\AgdaSymbol{(}\AgdaBound{F}\AgdaSpace{}%
\AgdaBound{f}\AgdaSymbol{)}\AgdaSpace{}%
\AgdaBound{e}\<%
\end{code}
}

\begin{code}[hide]%
\>[0]\AgdaFunction{list-map}\AgdaSpace{}%
\AgdaSymbol{:}\AgdaSpace{}%
\AgdaFunction{Map}\AgdaSpace{}%
\AgdaFunction{List}\<%
\\
\>[0]\AgdaFunction{nest-map}\AgdaSpace{}%
\AgdaSymbol{:}\AgdaSpace{}%
\AgdaFunction{Map}\AgdaSpace{}%
\AgdaFunction{Nest}\<%
\\
\>[0]\AgdaFunction{bush-map}\AgdaSpace{}%
\AgdaSymbol{:}\AgdaSpace{}%
\AgdaFunction{Map}\AgdaSpace{}%
\AgdaFunction{Bush}\<%
\\
\>[0]\AgdaFunction{maybe-map}\AgdaSpace{}%
\AgdaSymbol{:}\AgdaSpace{}%
\AgdaFunction{Map}\AgdaSpace{}%
\AgdaFunction{Maybe}\<%
\end{code}

\newcommand{\tuplesmap}{
\begin{code}%
\>[0]\AgdaFunction{tuple-map}\AgdaSpace{}%
\AgdaSymbol{:}\AgdaSpace{}%
\AgdaSymbol{∀}\AgdaSpace{}%
\AgdaBound{n}\AgdaSpace{}%
\AgdaSymbol{→}\AgdaSpace{}%
\AgdaFunction{Map}\AgdaSpace{}%
\AgdaSymbol{(}\AgdaFunction{Tuple}\AgdaSpace{}%
\AgdaBound{n}\AgdaSymbol{)}\<%
\\
\>[0]\AgdaFunction{tuple-map}\AgdaSpace{}%
\AgdaInductiveConstructor{zero}\AgdaSpace{}%
\AgdaSymbol{=}\AgdaSpace{}%
\AgdaFunction{id}\<%
\\
\>[0]\AgdaFunction{tuple-map}\AgdaSpace{}%
\AgdaSymbol{(}\AgdaInductiveConstructor{suc}\AgdaSpace{}%
\AgdaBound{n}\AgdaSymbol{)}\AgdaSpace{}%
\AgdaBound{f}\AgdaSpace{}%
\AgdaSymbol{(}\AgdaBound{a}\AgdaSpace{}%
\AgdaOperator{\AgdaInductiveConstructor{,}}\AgdaSpace{}%
\AgdaBound{ta}\AgdaSymbol{)}\AgdaSpace{}%
\AgdaSymbol{=}\AgdaSpace{}%
\AgdaSymbol{(}\AgdaBound{f}\AgdaSpace{}%
\AgdaBound{a}\AgdaSpace{}%
\AgdaOperator{\AgdaInductiveConstructor{,}}\AgdaSpace{}%
\AgdaFunction{tuple-map}\AgdaSpace{}%
\AgdaBound{n}\AgdaSpace{}%
\AgdaBound{f}\AgdaSpace{}%
\AgdaBound{ta}\AgdaSymbol{)}\<%
\end{code}
}

\newcommand{\lndtmaps}{
\begin{multicols}{2}
\begin{code}%
\>[0]\AgdaFunction{list-map}\AgdaSpace{}%
\AgdaSymbol{=}\AgdaSpace{}%
\AgdaFunction{lndt-map}\AgdaSpace{}%
\AgdaSymbol{(}\AgdaFunction{tuple-map}\AgdaSpace{}%
\AgdaNumber{0}\AgdaSymbol{)}\<%
\\
\>[0]\AgdaFunction{nest-map}\AgdaSpace{}%
\AgdaSymbol{=}\AgdaSpace{}%
\AgdaFunction{lndt-map}\AgdaSpace{}%
\AgdaSymbol{(}\AgdaFunction{tuple-map}\AgdaSpace{}%
\AgdaNumber{1}\AgdaSymbol{)}\<%
\end{code}
\columnbreak
\begin{code}[hide]%
\>[0]\AgdaSymbol{\{-\#}\AgdaSpace{}%
\AgdaKeyword{TERMINATING}\AgdaSpace{}%
\AgdaSymbol{\#-\}}\<%
\end{code}
\begin{code}%
\>[0]\AgdaFunction{bush-map}\AgdaSpace{}%
\AgdaSymbol{=}\AgdaSpace{}%
\AgdaFunction{lndt-map}\AgdaSpace{}%
\AgdaFunction{bush-map}\<%
\\
\>[0]\AgdaFunction{maybe-map}\AgdaSpace{}%
\AgdaSymbol{=}\AgdaSpace{}%
\AgdaFunction{lndt-map}\AgdaSpace{}%
\AgdaSymbol{(λ}\AgdaSpace{}%
\AgdaBound{\AgdaUnderscore{}}\AgdaSpace{}%
\AgdaSymbol{())}\<%
\end{code}
\end{multicols}
}

\begin{code}[hide]%
\>[0]\AgdaKeyword{infixr}\AgdaSpace{}%
\AgdaNumber{10}\AgdaSpace{}%
\AgdaOperator{\AgdaInductiveConstructor{\AgdaUnderscore{}∷\AgdaUnderscore{}}}\<%
\end{code}

\newcommand{\firstexamplelist}{
\begin{code}%
\>[0]\AgdaFunction{list-map-example}\AgdaSpace{}%
\AgdaSymbol{:}\AgdaSpace{}%
\AgdaFunction{list-map}\AgdaSpace{}%
\AgdaInductiveConstructor{suc}\AgdaSpace{}%
\AgdaSymbol{(}\AgdaNumber{3}\AgdaSpace{}%
\AgdaOperator{\AgdaInductiveConstructor{∷}}\AgdaSpace{}%
\AgdaNumber{4}\AgdaSpace{}%
\AgdaOperator{\AgdaInductiveConstructor{∷}}\AgdaSpace{}%
\AgdaNumber{2}\AgdaSpace{}%
\AgdaOperator{\AgdaInductiveConstructor{∷}}\AgdaSpace{}%
\AgdaNumber{6}\AgdaSpace{}%
\AgdaOperator{\AgdaInductiveConstructor{∷}}\AgdaSpace{}%
\AgdaInductiveConstructor{[]}\AgdaSymbol{)}\AgdaSpace{}%
\AgdaOperator{\AgdaDatatype{≡}}\AgdaSpace{}%
\AgdaNumber{4}\AgdaSpace{}%
\AgdaOperator{\AgdaInductiveConstructor{∷}}\AgdaSpace{}%
\AgdaNumber{5}\AgdaSpace{}%
\AgdaOperator{\AgdaInductiveConstructor{∷}}\AgdaSpace{}%
\AgdaNumber{3}\AgdaSpace{}%
\AgdaOperator{\AgdaInductiveConstructor{∷}}\AgdaSpace{}%
\AgdaNumber{7}\AgdaSpace{}%
\AgdaOperator{\AgdaInductiveConstructor{∷}}\AgdaSpace{}%
\AgdaInductiveConstructor{[]}\<%
\\
\>[0]\AgdaFunction{list-map-example}\AgdaSpace{}%
\AgdaSymbol{=}\AgdaSpace{}%
\AgdaInductiveConstructor{refl}\<%
\end{code}
}

\newcommand{\firstexamplebush}{
\begin{code}%
\>[0]\AgdaFunction{bush-map-example}\AgdaSpace{}%
\AgdaSymbol{:}\AgdaSpace{}%
\AgdaFunction{bush-map}\AgdaSpace{}%
\AgdaSymbol{(}\AgdaOperator{\AgdaPrimitive{\AgdaUnderscore{}*}}\AgdaSpace{}%
\AgdaNumber{2}\AgdaSymbol{)}\AgdaSpace{}%
\AgdaSymbol{(}\AgdaNumber{3}\AgdaSpace{}%
\AgdaOperator{\AgdaInductiveConstructor{∷}}\AgdaSpace{}%
\AgdaSymbol{(}\AgdaNumber{4}\AgdaSpace{}%
\AgdaOperator{\AgdaInductiveConstructor{∷}}\AgdaSpace{}%
\AgdaInductiveConstructor{[]}\AgdaSymbol{)}\AgdaSpace{}%
\AgdaOperator{\AgdaInductiveConstructor{∷}}\AgdaSpace{}%
\AgdaInductiveConstructor{[]}\AgdaSymbol{)}\AgdaSpace{}%
\AgdaOperator{\AgdaDatatype{≡}}\AgdaSpace{}%
\AgdaSymbol{(}\AgdaNumber{6}\AgdaSpace{}%
\AgdaOperator{\AgdaInductiveConstructor{∷}}\AgdaSpace{}%
\AgdaSymbol{(}\AgdaNumber{8}\AgdaSpace{}%
\AgdaOperator{\AgdaInductiveConstructor{∷}}\AgdaSpace{}%
\AgdaInductiveConstructor{[]}\AgdaSymbol{)}\AgdaSpace{}%
\AgdaOperator{\AgdaInductiveConstructor{∷}}\AgdaSpace{}%
\AgdaInductiveConstructor{[]}\AgdaSymbol{)}\<%
\\
\>[0]\AgdaFunction{bush-map-example}\AgdaSpace{}%
\AgdaSymbol{=}\AgdaSpace{}%
\AgdaInductiveConstructor{refl}\<%
\end{code}
}

\newcommand{\fold}{
\begin{code}%
\>[0]\AgdaFunction{Fold}\AgdaSpace{}%
\AgdaSymbol{:}\AgdaSpace{}%
\AgdaFunction{TT}\AgdaSpace{}%
\AgdaSymbol{→}\AgdaSpace{}%
\AgdaPrimitive{Setω}\<%
\\
\>[0]\AgdaFunction{Fold}\AgdaSpace{}%
\AgdaBound{F}\AgdaSpace{}%
\AgdaSymbol{=}\AgdaSpace{}%
\AgdaSymbol{∀}\AgdaSpace{}%
\AgdaSymbol{\{}\AgdaBound{a}\AgdaSpace{}%
\AgdaBound{b}\AgdaSymbol{\}}\AgdaSpace{}%
\AgdaSymbol{\{}\AgdaBound{A}\AgdaSpace{}%
\AgdaSymbol{:}\AgdaSpace{}%
\AgdaPrimitive{Set}\AgdaSpace{}%
\AgdaBound{a}\AgdaSymbol{\}}\AgdaSpace{}%
\AgdaSymbol{\{}\AgdaBound{B}\AgdaSpace{}%
\AgdaSymbol{:}\AgdaSpace{}%
\AgdaPrimitive{Set}\AgdaSpace{}%
\AgdaBound{b}\AgdaSymbol{\}}\AgdaSpace{}%
\AgdaSymbol{→}\AgdaSpace{}%
\AgdaSymbol{(}\AgdaBound{B}\AgdaSpace{}%
\AgdaSymbol{→}\AgdaSpace{}%
\AgdaBound{A}\AgdaSpace{}%
\AgdaSymbol{→}\AgdaSpace{}%
\AgdaBound{B}\AgdaSymbol{)}\AgdaSpace{}%
\AgdaSymbol{→}\AgdaSpace{}%
\AgdaBound{B}\AgdaSpace{}%
\AgdaSymbol{→}\AgdaSpace{}%
\AgdaBound{F}\AgdaSpace{}%
\AgdaBound{A}\AgdaSpace{}%
\AgdaSymbol{→}\AgdaSpace{}%
\AgdaBound{B}\<%
\end{code}
}

\newcommand{\folds}{
\begin{code}%
\>[0]\AgdaFunction{lndt-foldl}\AgdaSpace{}%
\AgdaSymbol{:}\AgdaSpace{}%
\AgdaSymbol{∀}\AgdaSpace{}%
\AgdaSymbol{\{}\AgdaBound{F}\AgdaSpace{}%
\AgdaSymbol{:}\AgdaSpace{}%
\AgdaFunction{TT}\AgdaSymbol{\}}\AgdaSpace{}%
\AgdaSymbol{→}\AgdaSpace{}%
\AgdaFunction{Fold}\AgdaSpace{}%
\AgdaBound{F}\AgdaSpace{}%
\AgdaSymbol{→}\AgdaSpace{}%
\AgdaFunction{Fold}\AgdaSpace{}%
\AgdaSymbol{(}\AgdaDatatype{LNDT}\AgdaSpace{}%
\AgdaBound{F}\AgdaSymbol{)}\<%
\\
\>[0]\AgdaFunction{lndt-foldl}\AgdaSpace{}%
\AgdaSymbol{\AgdaUnderscore{}}\AgdaSpace{}%
\AgdaSymbol{\AgdaUnderscore{}}\AgdaSpace{}%
\AgdaBound{b}\AgdaSpace{}%
\AgdaInductiveConstructor{[]}\AgdaSpace{}%
\AgdaSymbol{=}\AgdaSpace{}%
\AgdaBound{b}\<%
\\
\>[0]\AgdaFunction{lndt-foldl}\AgdaSpace{}%
\AgdaBound{foldl}\AgdaSpace{}%
\AgdaBound{f}\AgdaSpace{}%
\AgdaBound{b}\AgdaSpace{}%
\AgdaSymbol{(}\AgdaBound{x}\AgdaSpace{}%
\AgdaOperator{\AgdaInductiveConstructor{∷}}\AgdaSpace{}%
\AgdaBound{e}\AgdaSymbol{)}\AgdaSpace{}%
\AgdaSymbol{=}\AgdaSpace{}%
\AgdaFunction{lndt-foldl}\AgdaSpace{}%
\AgdaBound{foldl}\AgdaSpace{}%
\AgdaSymbol{(}\AgdaBound{foldl}\AgdaSpace{}%
\AgdaBound{f}\AgdaSymbol{)}\AgdaSpace{}%
\AgdaSymbol{(}\AgdaBound{f}\AgdaSpace{}%
\AgdaBound{b}\AgdaSpace{}%
\AgdaBound{x}\AgdaSymbol{)}\AgdaSpace{}%
\AgdaBound{e}\<%
\end{code}
}

\newcommand{\foldtuples}{
\begin{code}%
\>[0]\AgdaFunction{tuple-foldl}\AgdaSpace{}%
\AgdaSymbol{:}\AgdaSpace{}%
\AgdaSymbol{∀}\AgdaSpace{}%
\AgdaBound{n}\AgdaSpace{}%
\AgdaSymbol{→}\AgdaSpace{}%
\AgdaFunction{Fold}\AgdaSpace{}%
\AgdaSymbol{(}\AgdaFunction{Tuple}\AgdaSpace{}%
\AgdaBound{n}\AgdaSymbol{)}\<%
\\
\>[0]\AgdaFunction{tuple-foldl}\AgdaSpace{}%
\AgdaInductiveConstructor{zero}\AgdaSpace{}%
\AgdaSymbol{=}\AgdaSpace{}%
\AgdaFunction{id}\<%
\\
\>[0]\AgdaFunction{tuple-foldl}\AgdaSpace{}%
\AgdaSymbol{(}\AgdaInductiveConstructor{suc}\AgdaSpace{}%
\AgdaBound{n}\AgdaSymbol{)}\AgdaSpace{}%
\AgdaBound{f}\AgdaSpace{}%
\AgdaBound{b₀}\AgdaSpace{}%
\AgdaSymbol{(}\AgdaBound{a}\AgdaSpace{}%
\AgdaOperator{\AgdaInductiveConstructor{,}}\AgdaSpace{}%
\AgdaBound{ta}\AgdaSymbol{)}\AgdaSpace{}%
\AgdaSymbol{=}\AgdaSpace{}%
\AgdaFunction{tuple-foldl}\AgdaSpace{}%
\AgdaBound{n}\AgdaSpace{}%
\AgdaBound{f}\AgdaSpace{}%
\AgdaSymbol{(}\AgdaBound{f}\AgdaSpace{}%
\AgdaBound{b₀}\AgdaSpace{}%
\AgdaBound{a}\AgdaSymbol{)}\AgdaSpace{}%
\AgdaBound{ta}\<%
\end{code}
}

\newcommand{\foldsinstances}{
\begin{multicols}{2}
\begin{code}%
\>[0]\AgdaFunction{nest-foldl}\AgdaSpace{}%
\AgdaSymbol{:}\AgdaSpace{}%
\AgdaFunction{Fold}\AgdaSpace{}%
\AgdaFunction{Nest}\<%
\\
\>[0]\AgdaFunction{nest-foldl}\AgdaSpace{}%
\AgdaSymbol{=}\AgdaSpace{}%
\AgdaFunction{lndt-foldl}\AgdaSpace{}%
\AgdaSymbol{(}\AgdaFunction{tuple-foldl}\AgdaSpace{}%
\AgdaNumber{1}\AgdaSymbol{)}\<%
\end{code}
\columnbreak
\begin{code}[hide]%
\>[0]\AgdaSymbol{\{-\#}\AgdaSpace{}%
\AgdaKeyword{TERMINATING}\AgdaSpace{}%
\AgdaSymbol{\#-\}}\<%
\end{code}
\begin{code}%
\>[0]\AgdaFunction{bush-foldl}\AgdaSpace{}%
\AgdaSymbol{:}\AgdaSpace{}%
\AgdaFunction{Fold}\AgdaSpace{}%
\AgdaFunction{Bush}\<%
\\
\>[0]\AgdaFunction{bush-foldl}\AgdaSpace{}%
\AgdaSymbol{=}\AgdaSpace{}%
\AgdaFunction{lndt-foldl}\AgdaSpace{}%
\AgdaFunction{bush-foldl}\<%
\end{code}
\end{multicols}
}

\newcommand{\foldsexampleone}{
\begin{code}%
\>[0]\AgdaFunction{foldl₀}\AgdaSpace{}%
\AgdaSymbol{:}\AgdaSpace{}%
\AgdaFunction{nest-foldl}\AgdaSpace{}%
\AgdaOperator{\AgdaFunction{\AgdaUnderscore{}++\AgdaUnderscore{}}}\AgdaSpace{}%
\AgdaString{""}\AgdaSpace{}%
\AgdaSymbol{(}\AgdaString{"a"}\AgdaSpace{}%
\AgdaOperator{\AgdaInductiveConstructor{∷}}\AgdaSpace{}%
\AgdaSymbol{(}\AgdaString{"r"}\AgdaSpace{}%
\AgdaOperator{\AgdaInductiveConstructor{,}}\AgdaSpace{}%
\AgdaString{"t"}\AgdaSymbol{)}\AgdaSpace{}%
\AgdaOperator{\AgdaInductiveConstructor{∷}}\AgdaSpace{}%
\AgdaSymbol{((}\AgdaString{"i"}\AgdaSpace{}%
\AgdaOperator{\AgdaInductiveConstructor{,}}\AgdaSpace{}%
\AgdaString{"c"}\AgdaSymbol{)}\AgdaSpace{}%
\AgdaOperator{\AgdaInductiveConstructor{,}}\AgdaSpace{}%
\AgdaSymbol{(}\AgdaString{"l"}\AgdaSpace{}%
\AgdaOperator{\AgdaInductiveConstructor{,}}\AgdaSpace{}%
\AgdaString{"e"}\AgdaSymbol{))}\AgdaSpace{}%
\AgdaOperator{\AgdaInductiveConstructor{∷}}\AgdaSpace{}%
\AgdaInductiveConstructor{[]}\AgdaSymbol{)}\AgdaSpace{}%
\AgdaOperator{\AgdaDatatype{≡}}\AgdaSpace{}%
\AgdaString{"article"}\<%
\\
\>[0]\AgdaFunction{foldl₀}\AgdaSpace{}%
\AgdaSymbol{=}\AgdaSpace{}%
\AgdaInductiveConstructor{refl}\<%
\end{code}
}
\newcommand{\foldsexampletwo}{
\begin{code}%
\>[0]\AgdaFunction{foldl₁}\AgdaSpace{}%
\AgdaSymbol{:}\AgdaSpace{}%
\AgdaFunction{bush-foldl}\AgdaSpace{}%
\AgdaOperator{\AgdaFunction{\AgdaUnderscore{}++\AgdaUnderscore{}}}\AgdaSpace{}%
\AgdaString{"m"}\AgdaSpace{}%
\AgdaSymbol{(}\AgdaString{"s"}\AgdaSpace{}%
\AgdaOperator{\AgdaInductiveConstructor{∷}}\AgdaSpace{}%
\AgdaSymbol{(}\AgdaString{"f"}\AgdaSpace{}%
\AgdaOperator{\AgdaInductiveConstructor{∷}}\AgdaSpace{}%
\AgdaInductiveConstructor{[]}\AgdaSymbol{)}\AgdaSpace{}%
\AgdaOperator{\AgdaInductiveConstructor{∷}}\AgdaSpace{}%
\AgdaSymbol{((}\AgdaString{"p"}\AgdaSpace{}%
\AgdaOperator{\AgdaInductiveConstructor{∷}}\AgdaSpace{}%
\AgdaInductiveConstructor{[]}\AgdaSymbol{)}\AgdaSpace{}%
\AgdaOperator{\AgdaInductiveConstructor{∷}}\AgdaSpace{}%
\AgdaInductiveConstructor{[]}\AgdaSymbol{)}\AgdaSpace{}%
\AgdaOperator{\AgdaInductiveConstructor{∷}}\AgdaSpace{}%
\AgdaInductiveConstructor{[]}\AgdaSymbol{)}\AgdaSpace{}%
\AgdaOperator{\AgdaDatatype{≡}}\AgdaSpace{}%
\AgdaString{"msfp"}\<%
\\
\>[0]\AgdaFunction{foldl₁}\AgdaSpace{}%
\AgdaSymbol{=}\AgdaSpace{}%
\AgdaInductiveConstructor{refl}\<%
\end{code}
}

\newcommand{\lndtlist}{
\begin{code}%
\>[0]\AgdaFunction{SquaredList}\AgdaSpace{}%
\AgdaSymbol{:}\AgdaSpace{}%
\AgdaFunction{TT}\<%
\\
\>[0]\AgdaFunction{SquaredList}\AgdaSpace{}%
\AgdaSymbol{=}\AgdaSpace{}%
\AgdaDatatype{LNDT}\AgdaSpace{}%
\AgdaFunction{List}\<%
\end{code}
}

\newcommand{\lndtlistexample}{
\begin{code}%
\>[0]\AgdaFunction{squared-list-example}\AgdaSpace{}%
\AgdaSymbol{:}\AgdaSpace{}%
\AgdaFunction{SquaredList}\AgdaSpace{}%
\AgdaDatatype{ℕ}\<%
\\
\>[0]\AgdaFunction{squared-list-example}\AgdaSpace{}%
\AgdaSymbol{=}\AgdaSpace{}%
\AgdaNumber{8}\AgdaSpace{}%
\AgdaOperator{\AgdaInductiveConstructor{∷}}\AgdaSpace{}%
\AgdaSymbol{(}\AgdaNumber{4}\AgdaSpace{}%
\AgdaOperator{\AgdaInductiveConstructor{∷}}\AgdaSpace{}%
\AgdaNumber{5}\AgdaSpace{}%
\AgdaOperator{\AgdaInductiveConstructor{∷}}\AgdaSpace{}%
\AgdaInductiveConstructor{[]}\AgdaSymbol{)}\AgdaSpace{}%
\AgdaOperator{\AgdaInductiveConstructor{∷}}\AgdaSpace{}%
\AgdaSymbol{((}\AgdaNumber{3}\AgdaSpace{}%
\AgdaOperator{\AgdaInductiveConstructor{∷}}\AgdaSpace{}%
\AgdaNumber{6}\AgdaSpace{}%
\AgdaOperator{\AgdaInductiveConstructor{∷}}\AgdaSpace{}%
\AgdaInductiveConstructor{[]}\AgdaSymbol{)}\AgdaSpace{}%
\AgdaOperator{\AgdaInductiveConstructor{∷}}\AgdaSpace{}%
\AgdaSymbol{(}\AgdaNumber{7}\AgdaSpace{}%
\AgdaOperator{\AgdaInductiveConstructor{∷}}\AgdaSpace{}%
\AgdaNumber{1}\AgdaSpace{}%
\AgdaOperator{\AgdaInductiveConstructor{∷}}\AgdaSpace{}%
\AgdaNumber{8}\AgdaSpace{}%
\AgdaOperator{\AgdaInductiveConstructor{∷}}\AgdaSpace{}%
\AgdaInductiveConstructor{[]}\AgdaSymbol{)}\AgdaSpace{}%
\AgdaOperator{\AgdaInductiveConstructor{∷}}\AgdaSpace{}%
\AgdaInductiveConstructor{[]}\AgdaSymbol{)}\AgdaSpace{}%
\AgdaOperator{\AgdaInductiveConstructor{∷}}\AgdaSpace{}%
\AgdaInductiveConstructor{[]}\<%
\end{code}
}

\newcommand{\lndtlistmap}{
\begin{code}%
\>[0]\AgdaFunction{squared-list-map}\AgdaSpace{}%
\AgdaSymbol{:}\AgdaSpace{}%
\AgdaFunction{Map}\AgdaSpace{}%
\AgdaFunction{SquaredList}\<%
\\
\>[0]\AgdaFunction{squared-list-map}\AgdaSpace{}%
\AgdaSymbol{=}\AgdaSpace{}%
\AgdaFunction{lndt-map}\AgdaSpace{}%
\AgdaFunction{list-map}\<%
\end{code}
}

\newcommand{\lndtlistmapexample}{
\begin{code}%
\>[0]\AgdaFunction{squared-list-map-example}\AgdaSpace{}%
\AgdaSymbol{:}\AgdaSpace{}%
\AgdaFunction{squared-list-map}\AgdaSpace{}%
\AgdaSymbol{(}\AgdaOperator{\AgdaPrimitive{\AgdaUnderscore{}*}}\AgdaSpace{}%
\AgdaNumber{2}\AgdaSymbol{)}\AgdaSpace{}%
\AgdaFunction{squared-list-example}\<%
\\
\>[0][@{}l@{\AgdaIndent{0}}]%
\>[2]\AgdaOperator{\AgdaDatatype{≡}}\AgdaSpace{}%
\AgdaNumber{16}\AgdaSpace{}%
\AgdaOperator{\AgdaInductiveConstructor{∷}}\AgdaSpace{}%
\AgdaSymbol{(}\AgdaNumber{8}\AgdaSpace{}%
\AgdaOperator{\AgdaInductiveConstructor{∷}}\AgdaSpace{}%
\AgdaNumber{10}\AgdaSpace{}%
\AgdaOperator{\AgdaInductiveConstructor{∷}}\AgdaSpace{}%
\AgdaInductiveConstructor{[]}\AgdaSymbol{)}\AgdaSpace{}%
\AgdaOperator{\AgdaInductiveConstructor{∷}}\AgdaSpace{}%
\AgdaSymbol{((}\AgdaNumber{6}\AgdaSpace{}%
\AgdaOperator{\AgdaInductiveConstructor{∷}}\AgdaSpace{}%
\AgdaNumber{12}\AgdaSpace{}%
\AgdaOperator{\AgdaInductiveConstructor{∷}}\AgdaSpace{}%
\AgdaInductiveConstructor{[]}\AgdaSymbol{)}\AgdaSpace{}%
\AgdaOperator{\AgdaInductiveConstructor{∷}}\AgdaSpace{}%
\AgdaSymbol{(}\AgdaNumber{14}\AgdaSpace{}%
\AgdaOperator{\AgdaInductiveConstructor{∷}}\AgdaSpace{}%
\AgdaNumber{2}\AgdaSpace{}%
\AgdaOperator{\AgdaInductiveConstructor{∷}}\AgdaSpace{}%
\AgdaNumber{16}\AgdaSpace{}%
\AgdaOperator{\AgdaInductiveConstructor{∷}}\AgdaSpace{}%
\AgdaInductiveConstructor{[]}\AgdaSymbol{)}\AgdaSpace{}%
\AgdaOperator{\AgdaInductiveConstructor{∷}}\AgdaSpace{}%
\AgdaInductiveConstructor{[]}\AgdaSymbol{)}\AgdaSpace{}%
\AgdaOperator{\AgdaInductiveConstructor{∷}}\AgdaSpace{}%
\AgdaInductiveConstructor{[]}\<%
\\
\>[0]\AgdaFunction{squared-list-map-example}\AgdaSpace{}%
\AgdaSymbol{=}\AgdaSpace{}%
\AgdaInductiveConstructor{refl}\<%
\end{code}
}

\newcommand{\transpred}{
\begin{code}%
\>[0]\AgdaFunction{TransPred}\AgdaSpace{}%
\AgdaSymbol{:}\AgdaSpace{}%
\AgdaFunction{TT}\AgdaSpace{}%
\AgdaSymbol{→}\AgdaSpace{}%
\AgdaPrimitive{Setω}\<%
\\
\>[0]\AgdaFunction{TransPred}\AgdaSpace{}%
\AgdaBound{F}\AgdaSpace{}%
\AgdaSymbol{=}\AgdaSpace{}%
\AgdaSymbol{∀}\AgdaSpace{}%
\AgdaSymbol{\{}\AgdaBound{a}\AgdaSpace{}%
\AgdaBound{b}\AgdaSymbol{\}}\AgdaSpace{}%
\AgdaSymbol{\{}\AgdaBound{A}\AgdaSpace{}%
\AgdaSymbol{:}\AgdaSpace{}%
\AgdaPrimitive{Set}\AgdaSpace{}%
\AgdaBound{a}\AgdaSymbol{\}}\AgdaSpace{}%
\AgdaSymbol{→}\AgdaSpace{}%
\AgdaFunction{Pred}\AgdaSpace{}%
\AgdaBound{A}\AgdaSpace{}%
\AgdaBound{b}\AgdaSpace{}%
\AgdaSymbol{→}\AgdaSpace{}%
\AgdaFunction{Pred}\AgdaSpace{}%
\AgdaSymbol{(}\AgdaBound{F}\AgdaSpace{}%
\AgdaBound{A}\AgdaSymbol{)}\AgdaSpace{}%
\AgdaBound{b}\<%
\end{code}
}

\newcommand{\transdec}{
\begin{code}%
\>[0]\AgdaFunction{TransDec}\AgdaSpace{}%
\AgdaSymbol{:}\AgdaSpace{}%
\AgdaSymbol{∀}\AgdaSpace{}%
\AgdaSymbol{\{}\AgdaBound{F}\AgdaSpace{}%
\AgdaSymbol{:}\AgdaSpace{}%
\AgdaFunction{TT}\AgdaSymbol{\}}\AgdaSpace{}%
\AgdaSymbol{→}\AgdaSpace{}%
\AgdaFunction{TransPred}\AgdaSpace{}%
\AgdaBound{F}\AgdaSpace{}%
\AgdaSymbol{→}\AgdaSpace{}%
\AgdaPrimitive{Setω}\<%
\\
\>[0]\AgdaFunction{TransDec}\AgdaSpace{}%
\AgdaBound{TP}\AgdaSpace{}%
\AgdaSymbol{=}\AgdaSpace{}%
\AgdaSymbol{∀}\AgdaSpace{}%
\AgdaSymbol{\{}\AgdaBound{a}\AgdaSpace{}%
\AgdaBound{b}\AgdaSymbol{\}}\AgdaSpace{}%
\AgdaSymbol{\{}\AgdaBound{A}\AgdaSpace{}%
\AgdaSymbol{:}\AgdaSpace{}%
\AgdaPrimitive{Set}\AgdaSpace{}%
\AgdaBound{a}\AgdaSymbol{\}}\AgdaSpace{}%
\AgdaSymbol{\{}\AgdaBound{P}\AgdaSpace{}%
\AgdaSymbol{:}\AgdaSpace{}%
\AgdaFunction{Pred}\AgdaSpace{}%
\AgdaBound{A}\AgdaSpace{}%
\AgdaBound{b}\AgdaSymbol{\}}\AgdaSpace{}%
\AgdaSymbol{→}\AgdaSpace{}%
\AgdaFunction{Decidable}\AgdaSpace{}%
\AgdaBound{P}\AgdaSpace{}%
\AgdaSymbol{→}\AgdaSpace{}%
\AgdaFunction{Decidable}\AgdaSpace{}%
\AgdaSymbol{(}\AgdaBound{TP}\AgdaSpace{}%
\AgdaBound{P}\AgdaSymbol{)}\<%
\end{code}
}

\newcommand{\transany}{
\begin{code}%
\>[0]\AgdaKeyword{data}\AgdaSpace{}%
\AgdaDatatype{lndt-any}\AgdaSpace{}%
\AgdaSymbol{\{}\AgdaBound{F}\AgdaSpace{}%
\AgdaSymbol{:}\AgdaSpace{}%
\AgdaFunction{TT}\AgdaSymbol{\}}\AgdaSpace{}%
\AgdaSymbol{(}\AgdaBound{T}\AgdaSpace{}%
\AgdaSymbol{:}\AgdaSpace{}%
\AgdaFunction{TransPred}\AgdaSpace{}%
\AgdaBound{F}\AgdaSymbol{)}\AgdaSpace{}%
\AgdaSymbol{\{}\AgdaBound{a}\AgdaSpace{}%
\AgdaBound{b}\AgdaSymbol{\}}\AgdaSpace{}%
\AgdaSymbol{\{}\AgdaBound{A}\AgdaSpace{}%
\AgdaSymbol{:}\AgdaSpace{}%
\AgdaPrimitive{Set}\AgdaSpace{}%
\AgdaBound{a}\AgdaSymbol{\}}\AgdaSpace{}%
\AgdaSymbol{(}\AgdaBound{P}\AgdaSpace{}%
\AgdaSymbol{:}\AgdaSpace{}%
\AgdaFunction{Pred}\AgdaSpace{}%
\AgdaBound{A}\AgdaSpace{}%
\AgdaBound{b}\AgdaSymbol{)}\AgdaSpace{}%
\AgdaSymbol{:}\<%
\\
\>[0][@{}l@{\AgdaIndent{0}}]%
\>[2]\AgdaFunction{Pred}\AgdaSpace{}%
\AgdaSymbol{(}\AgdaDatatype{LNDT}\AgdaSpace{}%
\AgdaBound{F}\AgdaSpace{}%
\AgdaBound{A}\AgdaSymbol{)}\AgdaSpace{}%
\AgdaBound{b}\AgdaSpace{}%
\AgdaKeyword{where}\<%
\\
\>[2][@{}l@{\AgdaIndent{0}}]%
\>[4]\AgdaInductiveConstructor{here}\AgdaSpace{}%
\AgdaSymbol{:}\AgdaSpace{}%
\AgdaSymbol{∀}\AgdaSpace{}%
\AgdaSymbol{\{}\AgdaBound{a}\AgdaSpace{}%
\AgdaBound{x}\AgdaSymbol{\}}\AgdaSpace{}%
\AgdaSymbol{→}\AgdaSpace{}%
\AgdaBound{P}\AgdaSpace{}%
\AgdaBound{a}\AgdaSpace{}%
\AgdaSymbol{→}\AgdaSpace{}%
\AgdaDatatype{lndt-any}\AgdaSpace{}%
\AgdaBound{T}\AgdaSpace{}%
\AgdaBound{P}\AgdaSpace{}%
\AgdaSymbol{(}\AgdaBound{a}\AgdaSpace{}%
\AgdaOperator{\AgdaInductiveConstructor{∷}}\AgdaSpace{}%
\AgdaBound{x}\AgdaSymbol{)}\<%
\\
\>[4]\AgdaInductiveConstructor{there}\AgdaSpace{}%
\AgdaSymbol{:}\AgdaSpace{}%
\AgdaSymbol{∀}\AgdaSpace{}%
\AgdaSymbol{\{}\AgdaBound{a}\AgdaSpace{}%
\AgdaBound{x}\AgdaSymbol{\}}\AgdaSpace{}%
\AgdaSymbol{→}\AgdaSpace{}%
\AgdaDatatype{lndt-any}\AgdaSpace{}%
\AgdaBound{T}\AgdaSpace{}%
\AgdaSymbol{(}\AgdaBound{T}\AgdaSpace{}%
\AgdaBound{P}\AgdaSymbol{)}\AgdaSpace{}%
\AgdaBound{x}\AgdaSpace{}%
\AgdaSymbol{→}\AgdaSpace{}%
\AgdaDatatype{lndt-any}\AgdaSpace{}%
\AgdaBound{T}\AgdaSpace{}%
\AgdaBound{P}\AgdaSpace{}%
\AgdaSymbol{(}\AgdaBound{a}\AgdaSpace{}%
\AgdaOperator{\AgdaInductiveConstructor{∷}}\AgdaSpace{}%
\AgdaBound{x}\AgdaSymbol{)}\<%
\end{code}
}

\newcommand{\tupleany}{
\begin{code}%
\>[0]\AgdaFunction{tuple-any}\AgdaSpace{}%
\AgdaSymbol{:}\AgdaSpace{}%
\AgdaSymbol{∀}\AgdaSpace{}%
\AgdaBound{n}\AgdaSpace{}%
\AgdaSymbol{→}\AgdaSpace{}%
\AgdaFunction{TransPred}\AgdaSpace{}%
\AgdaSymbol{(}\AgdaFunction{Tuple}\AgdaSpace{}%
\AgdaBound{n}\AgdaSymbol{)}\<%
\\
\>[0]\AgdaFunction{tuple-any}\AgdaSpace{}%
\AgdaInductiveConstructor{zero}\AgdaSpace{}%
\AgdaSymbol{=}\AgdaSpace{}%
\AgdaFunction{id}\<%
\\
\>[0]\AgdaFunction{tuple-any}\AgdaSpace{}%
\AgdaSymbol{(}\AgdaInductiveConstructor{suc}\AgdaSpace{}%
\AgdaBound{n}\AgdaSymbol{)}\AgdaSpace{}%
\AgdaBound{P}\AgdaSpace{}%
\AgdaSymbol{(}\AgdaBound{a}\AgdaSpace{}%
\AgdaOperator{\AgdaInductiveConstructor{,}}\AgdaSpace{}%
\AgdaBound{t}\AgdaSymbol{)}\AgdaSpace{}%
\AgdaSymbol{=}\AgdaSpace{}%
\AgdaBound{P}\AgdaSpace{}%
\AgdaBound{a}\AgdaSpace{}%
\AgdaOperator{\AgdaDatatype{⊎}}\AgdaSpace{}%
\AgdaFunction{tuple-any}\AgdaSpace{}%
\AgdaBound{n}\AgdaSpace{}%
\AgdaBound{P}\AgdaSpace{}%
\AgdaBound{t}\<%
\end{code}
}

\newcommand{\anyinstances}{
\begin{multicols}{2}
\begin{code}%
\>[0]\AgdaFunction{nest-any}\AgdaSpace{}%
\AgdaSymbol{:}\AgdaSpace{}%
\AgdaFunction{TransPred}\AgdaSpace{}%
\AgdaFunction{Nest}\<%
\\
\>[0]\AgdaFunction{nest-any}\AgdaSpace{}%
\AgdaSymbol{=}\AgdaSpace{}%
\AgdaDatatype{lndt-any}\AgdaSpace{}%
\AgdaSymbol{(}\AgdaFunction{tuple-any}\AgdaSpace{}%
\AgdaNumber{1}\AgdaSymbol{)}\<%
\end{code}
\columnbreak
\begin{code}[hide]%
\>[0]\AgdaSymbol{\{-\#}\AgdaSpace{}%
\AgdaKeyword{TERMINATING}\AgdaSpace{}%
\AgdaSymbol{\#-\}}\<%
\end{code}
\begin{code}%
\>[0]\AgdaFunction{bush-any}\AgdaSpace{}%
\AgdaSymbol{:}\AgdaSpace{}%
\AgdaFunction{TransPred}\AgdaSpace{}%
\AgdaFunction{Bush}\<%
\\
\>[0]\AgdaFunction{bush-any}\AgdaSpace{}%
\AgdaSymbol{=}\AgdaSpace{}%
\AgdaDatatype{lndt-any}\AgdaSpace{}%
\AgdaFunction{bush-any}\<%
\end{code}
\end{multicols}
}

\newcommand{\anybushexample}{
\begin{code}%
\>[0]\AgdaFunction{bush-any-example}\AgdaSpace{}%
\AgdaSymbol{:}\AgdaSpace{}%
\AgdaFunction{bush-any}\AgdaSpace{}%
\AgdaSymbol{(}\AgdaOperator{\AgdaDatatype{\AgdaUnderscore{}≡}}\AgdaSpace{}%
\AgdaNumber{10}\AgdaSymbol{)}\AgdaSpace{}%
\AgdaSymbol{(}\AgdaNumber{3}\AgdaSpace{}%
\AgdaOperator{\AgdaInductiveConstructor{∷}}\AgdaSpace{}%
\AgdaInductiveConstructor{[]}\AgdaSpace{}%
\AgdaOperator{\AgdaInductiveConstructor{∷}}\AgdaSpace{}%
\AgdaSymbol{((}\AgdaNumber{4}\AgdaSpace{}%
\AgdaOperator{\AgdaInductiveConstructor{∷}}\AgdaSpace{}%
\AgdaSymbol{(}\AgdaNumber{7}\AgdaSpace{}%
\AgdaOperator{\AgdaInductiveConstructor{∷}}\AgdaSpace{}%
\AgdaInductiveConstructor{[]}\AgdaSymbol{)}\AgdaSpace{}%
\AgdaOperator{\AgdaInductiveConstructor{∷}}\AgdaSpace{}%
\AgdaInductiveConstructor{[]}\AgdaSymbol{)}\AgdaSpace{}%
\AgdaOperator{\AgdaInductiveConstructor{∷}}\AgdaSpace{}%
\AgdaSymbol{((}\AgdaNumber{10}\AgdaSpace{}%
\AgdaOperator{\AgdaInductiveConstructor{∷}}\AgdaSpace{}%
\AgdaInductiveConstructor{[]}\AgdaSymbol{)}\AgdaSpace{}%
\AgdaOperator{\AgdaInductiveConstructor{∷}}\AgdaSpace{}%
\AgdaInductiveConstructor{[]}\AgdaSymbol{)}\AgdaSpace{}%
\AgdaOperator{\AgdaInductiveConstructor{∷}}\AgdaSpace{}%
\AgdaInductiveConstructor{[]}\AgdaSymbol{)}\AgdaSpace{}%
\AgdaOperator{\AgdaInductiveConstructor{∷}}\AgdaSpace{}%
\AgdaInductiveConstructor{[]}\AgdaSymbol{)}\<%
\\
\>[0]\AgdaFunction{bush-any-example}\AgdaSpace{}%
\AgdaSymbol{=}\AgdaSpace{}%
\AgdaInductiveConstructor{there}\AgdaSpace{}%
\AgdaSymbol{(}\AgdaInductiveConstructor{there}\AgdaSpace{}%
\AgdaSymbol{(}\AgdaInductiveConstructor{here}\AgdaSpace{}%
\AgdaSymbol{(}\AgdaInductiveConstructor{there}\AgdaSpace{}%
\AgdaSymbol{(}\AgdaInductiveConstructor{here}\AgdaSpace{}%
\AgdaSymbol{(}\AgdaInductiveConstructor{here}\AgdaSpace{}%
\AgdaSymbol{(}\AgdaInductiveConstructor{here}\AgdaSpace{}%
\AgdaInductiveConstructor{refl}\AgdaSymbol{))))))}\<%
\end{code}
}

\newcommand{\lndtdecany}{
\begin{code}%
\>[0]\AgdaFunction{lndt-dec-any}\AgdaSpace{}%
\AgdaSymbol{:}\AgdaSpace{}%
\AgdaSymbol{∀}\AgdaSpace{}%
\AgdaSymbol{\{}\AgdaBound{F}\AgdaSpace{}%
\AgdaSymbol{:}\AgdaSpace{}%
\AgdaFunction{TT}\AgdaSymbol{\}}\AgdaSpace{}%
\AgdaSymbol{\{}\AgdaBound{T}\AgdaSpace{}%
\AgdaSymbol{:}\AgdaSpace{}%
\AgdaFunction{TransPred}\AgdaSpace{}%
\AgdaBound{F}\AgdaSymbol{\}}\AgdaSpace{}%
\AgdaSymbol{→}\AgdaSpace{}%
\AgdaFunction{TransDec}\AgdaSpace{}%
\AgdaBound{T}\AgdaSpace{}%
\AgdaSymbol{→}\AgdaSpace{}%
\AgdaFunction{TransDec}\AgdaSpace{}%
\AgdaSymbol{(}\AgdaDatatype{lndt-any}\AgdaSpace{}%
\AgdaBound{T}\AgdaSymbol{)}\<%
\end{code}
}

\begin{code}[hide]%
\>[0]\AgdaFunction{lndt-dec-any}\AgdaSpace{}%
\AgdaSymbol{\AgdaUnderscore{}}\AgdaSpace{}%
\AgdaSymbol{\AgdaUnderscore{}}\AgdaSpace{}%
\AgdaInductiveConstructor{[]}\AgdaSpace{}%
\AgdaSymbol{=}\AgdaSpace{}%
\AgdaInductiveConstructor{no}\AgdaSpace{}%
\AgdaSymbol{(λ}\AgdaSpace{}%
\AgdaSymbol{())}\<%
\\
\>[0]\AgdaFunction{lndt-dec-any}\AgdaSpace{}%
\AgdaBound{tdec}\AgdaSpace{}%
\AgdaBound{decP}\AgdaSpace{}%
\AgdaSymbol{(}\AgdaBound{x}\AgdaSpace{}%
\AgdaOperator{\AgdaInductiveConstructor{∷}}\AgdaSpace{}%
\AgdaBound{v}\AgdaSymbol{)}\AgdaSpace{}%
\AgdaKeyword{with}\AgdaSpace{}%
\AgdaBound{decP}\AgdaSpace{}%
\AgdaBound{x}\AgdaSpace{}%
\AgdaSymbol{|}\AgdaSpace{}%
\AgdaFunction{lndt-dec-any}\AgdaSpace{}%
\AgdaBound{tdec}\AgdaSpace{}%
\AgdaSymbol{(}\AgdaBound{tdec}\AgdaSpace{}%
\AgdaBound{decP}\AgdaSymbol{)}\AgdaSpace{}%
\AgdaBound{v}\<%
\\
\>[0]\AgdaSymbol{...}\AgdaSpace{}%
\AgdaSymbol{|}\AgdaSpace{}%
\AgdaInductiveConstructor{yes}\AgdaSpace{}%
\AgdaBound{p}%
\>[13]\AgdaSymbol{|}\AgdaSpace{}%
\AgdaSymbol{\AgdaUnderscore{}}%
\>[22]\AgdaSymbol{=}\AgdaSpace{}%
\AgdaInductiveConstructor{yes}\AgdaSpace{}%
\AgdaSymbol{(}\AgdaInductiveConstructor{here}\AgdaSpace{}%
\AgdaBound{p}\AgdaSymbol{)}\<%
\\
\>[0]\AgdaSymbol{...}\AgdaSpace{}%
\AgdaSymbol{|}\AgdaSpace{}%
\AgdaInductiveConstructor{no}%
\>[10]\AgdaSymbol{\AgdaUnderscore{}}%
\>[13]\AgdaSymbol{|}\AgdaSpace{}%
\AgdaInductiveConstructor{yes}\AgdaSpace{}%
\AgdaBound{q}%
\>[22]\AgdaSymbol{=}\AgdaSpace{}%
\AgdaInductiveConstructor{yes}\AgdaSpace{}%
\AgdaSymbol{(}\AgdaInductiveConstructor{there}\AgdaSpace{}%
\AgdaBound{q}\AgdaSymbol{)}\<%
\\
\>[0]\AgdaSymbol{...}\AgdaSpace{}%
\AgdaSymbol{|}\AgdaSpace{}%
\AgdaInductiveConstructor{no}%
\>[10]\AgdaBound{¬p}\AgdaSpace{}%
\AgdaSymbol{|}\AgdaSpace{}%
\AgdaInductiveConstructor{no}%
\>[19]\AgdaBound{¬q}\AgdaSpace{}%
\AgdaSymbol{=}\AgdaSpace{}%
\AgdaInductiveConstructor{no}\AgdaSpace{}%
\AgdaSymbol{(λ}\AgdaSpace{}%
\AgdaSymbol{\{(}\AgdaInductiveConstructor{here}\AgdaSpace{}%
\AgdaBound{p}\AgdaSymbol{)}\AgdaSpace{}%
\AgdaSymbol{→}\AgdaSpace{}%
\AgdaBound{¬p}\AgdaSpace{}%
\AgdaBound{p}\AgdaSpace{}%
\AgdaSymbol{;}\AgdaSpace{}%
\AgdaSymbol{(}\AgdaInductiveConstructor{there}\AgdaSpace{}%
\AgdaBound{q}\AgdaSymbol{)}\AgdaSpace{}%
\AgdaSymbol{→}\AgdaSpace{}%
\AgdaBound{¬q}\AgdaSpace{}%
\AgdaBound{q}\AgdaSymbol{\})}\<%
\end{code}

\newcommand{\mapcongruence}{
\begin{code}%
\>[0]\AgdaFunction{MapCongruence}\AgdaSpace{}%
\AgdaSymbol{:}\AgdaSpace{}%
\AgdaSymbol{∀}\AgdaSpace{}%
\AgdaSymbol{\{}\AgdaBound{F}\AgdaSpace{}%
\AgdaSymbol{:}\AgdaSpace{}%
\AgdaFunction{TT}\AgdaSymbol{\}}\AgdaSpace{}%
\AgdaSymbol{→}\AgdaSpace{}%
\AgdaFunction{Map}\AgdaSpace{}%
\AgdaBound{F}\AgdaSpace{}%
\AgdaSymbol{→}\AgdaSpace{}%
\AgdaPrimitive{Setω}\<%
\\
\>[0]\AgdaFunction{MapCongruence}\AgdaSpace{}%
\AgdaBound{map}\AgdaSpace{}%
\AgdaSymbol{=}\AgdaSpace{}%
\AgdaSymbol{∀}\AgdaSpace{}%
\AgdaSymbol{\{}\AgdaBound{a}\AgdaSpace{}%
\AgdaBound{b}\AgdaSymbol{\}}\AgdaSpace{}%
\AgdaSymbol{\{}\AgdaBound{A}\AgdaSpace{}%
\AgdaSymbol{:}\AgdaSpace{}%
\AgdaPrimitive{Set}\AgdaSpace{}%
\AgdaBound{a}\AgdaSymbol{\}}\AgdaSpace{}%
\AgdaSymbol{\{}\AgdaBound{B}\AgdaSpace{}%
\AgdaSymbol{:}\AgdaSpace{}%
\AgdaPrimitive{Set}\AgdaSpace{}%
\AgdaBound{b}\AgdaSymbol{\}}\AgdaSpace{}%
\AgdaSymbol{(}\AgdaBound{f}\AgdaSpace{}%
\AgdaBound{g}\AgdaSpace{}%
\AgdaSymbol{:}\AgdaSpace{}%
\AgdaBound{A}\AgdaSpace{}%
\AgdaSymbol{→}\AgdaSpace{}%
\AgdaBound{B}\AgdaSymbol{)}\AgdaSpace{}%
\AgdaSymbol{→}\<%
\\
\>[0][@{}l@{\AgdaIndent{0}}]%
\>[2]\AgdaSymbol{(∀}\AgdaSpace{}%
\AgdaBound{x}\AgdaSpace{}%
\AgdaSymbol{→}\AgdaSpace{}%
\AgdaBound{f}\AgdaSpace{}%
\AgdaBound{x}\AgdaSpace{}%
\AgdaOperator{\AgdaDatatype{≡}}\AgdaSpace{}%
\AgdaBound{g}\AgdaSpace{}%
\AgdaBound{x}\AgdaSymbol{)}\AgdaSpace{}%
\AgdaSymbol{→}\AgdaSpace{}%
\AgdaSymbol{(∀}\AgdaSpace{}%
\AgdaBound{x}\AgdaSpace{}%
\AgdaSymbol{→}\AgdaSpace{}%
\AgdaBound{map}\AgdaSpace{}%
\AgdaBound{f}\AgdaSpace{}%
\AgdaBound{x}\AgdaSpace{}%
\AgdaOperator{\AgdaDatatype{≡}}\AgdaSpace{}%
\AgdaBound{map}\AgdaSpace{}%
\AgdaBound{g}\AgdaSpace{}%
\AgdaBound{x}\AgdaSymbol{)}\<%
\end{code}
}

\newcommand{\mapcongproof}{
\begin{code}%
\>[0]\AgdaFunction{lndt-map-cong}\AgdaSpace{}%
\AgdaSymbol{:}\AgdaSpace{}%
\AgdaSymbol{∀}\AgdaSpace{}%
\AgdaSymbol{\{}\AgdaBound{F}\AgdaSpace{}%
\AgdaSymbol{:}\AgdaSpace{}%
\AgdaFunction{TT}\AgdaSymbol{\}}\AgdaSpace{}%
\AgdaSymbol{\{}\AgdaBound{map}\AgdaSpace{}%
\AgdaSymbol{:}\AgdaSpace{}%
\AgdaFunction{Map}\AgdaSpace{}%
\AgdaBound{F}\AgdaSymbol{\}}\<%
\\
\>[0][@{}l@{\AgdaIndent{0}}]%
\>[2]\AgdaSymbol{→}\AgdaSpace{}%
\AgdaFunction{MapCongruence}\AgdaSpace{}%
\AgdaBound{map}\AgdaSpace{}%
\AgdaSymbol{→}\AgdaSpace{}%
\AgdaFunction{MapCongruence}\AgdaSpace{}%
\AgdaSymbol{(}\AgdaFunction{lndt-map}\AgdaSpace{}%
\AgdaBound{map}\AgdaSymbol{)}\<%
\\
\>[0]\AgdaFunction{lndt-map-cong}\AgdaSpace{}%
\AgdaSymbol{\AgdaUnderscore{}}\AgdaSpace{}%
\AgdaSymbol{\AgdaUnderscore{}}\AgdaSpace{}%
\AgdaSymbol{\AgdaUnderscore{}}\AgdaSpace{}%
\AgdaSymbol{\AgdaUnderscore{}}\AgdaSpace{}%
\AgdaInductiveConstructor{[]}\AgdaSpace{}%
\AgdaSymbol{=}\AgdaSpace{}%
\AgdaInductiveConstructor{refl}\<%
\\
\>[0]\AgdaFunction{lndt-map-cong}\AgdaSpace{}%
\AgdaBound{cgMap}\AgdaSpace{}%
\AgdaBound{f}\AgdaSpace{}%
\AgdaBound{g}\AgdaSpace{}%
\AgdaBound{p}\AgdaSpace{}%
\AgdaSymbol{(}\AgdaBound{x}\AgdaSpace{}%
\AgdaOperator{\AgdaInductiveConstructor{∷}}\AgdaSpace{}%
\AgdaBound{v}\AgdaSymbol{)}\AgdaSpace{}%
\AgdaKeyword{rewrite}\AgdaSpace{}%
\AgdaBound{p}\AgdaSpace{}%
\AgdaBound{x}\AgdaSpace{}%
\AgdaSymbol{=}\<%
\\
\>[0][@{}l@{\AgdaIndent{0}}]%
\>[2]\AgdaFunction{cong}\AgdaSpace{}%
\AgdaSymbol{(}\AgdaBound{g}\AgdaSpace{}%
\AgdaBound{x}\AgdaSpace{}%
\AgdaOperator{\AgdaInductiveConstructor{∷\AgdaUnderscore{}}}\AgdaSymbol{)}\AgdaSpace{}%
\AgdaSymbol{(}\AgdaFunction{lndt-map-cong}\AgdaSpace{}%
\AgdaBound{cgMap}\AgdaSpace{}%
\AgdaSymbol{\AgdaUnderscore{}}\AgdaSpace{}%
\AgdaSymbol{\AgdaUnderscore{}}\AgdaSpace{}%
\AgdaSymbol{(}\AgdaBound{cgMap}\AgdaSpace{}%
\AgdaBound{f}\AgdaSpace{}%
\AgdaBound{g}\AgdaSpace{}%
\AgdaBound{p}\AgdaSymbol{)}\AgdaSpace{}%
\AgdaBound{v}\AgdaSymbol{)}\<%
\end{code}
}

\usepackage{breakurl}             % Not needed if you use pdflatex only.
\usepackage{underscore}           % Only needed if you use pdflatex.

\title{\libName: Towards a Formal Library on Spreadable Properties over Linked Nested Datatypes}

\author{Mathieu Montin
\institute{Université de Lorraine, Loria\\CNRS, Inria, France}
\email{mathieu.montin@loria.fr}
\and
Amélie Ledein
\institute{Université Paris-Saclay\\ENS Paris-Saclay, LMF\\CNRS, Inria, France}
\email{amelie.ledein@inria.fr}
\and
Catherine Dubois
\institute{ENSIIE, Samovar\\IP Paris, France}
\email{catherine.dubois@ensiie.fr}
}

\def\titlerunning{\libName: Towards a Formal Library on Spreadable Properties over Linked Nested Datatypes}
\def\authorrunning{M. Montin, A. Ledein and C. Dubois}

\ifthenelse{0 = 1}{
  \newcommand{\cath}[1]{\textcolor{Blue}{CD : #1}}
  \newcommand{\amelie}[1]{\textcolor{Red}{AL : #1}}
  \newcommand{\mathieu}[1]{\textcolor{Green}{MM : #1}}
}
{
  \newcommand{\mathieu}[1]{}
  \newcommand{\cath}[1]{}
  \newcommand{\amelie}[1]{}
}

\begin{document}
\maketitle

\begin{abstract}
Nested \datatypes have been widely studied in the past 25 years, both theoretically using category theory, and practically in programming languages such as \haskell. They consist in recursive polymorphic \datatypes where the type parameter changes throughout the recursion. They have a variety of applications such as modelling memory or modelling constraints over regular \datatypes without relying on dependent types. In this work, we focus on a specific subset of nested \datatypes which we call \textit{Linked Nested DataTypes} (\linear). We show that some usual \datatypes such has \textit{List} and \textit{Maybe}, as well as some well-known nested \datatypes such as \textit{Nest} and even \textit{Bush} can be built as various instances of \linear. We proceed by presenting \libName, a library, developed both in \agda and \coq, which focuses on the set of constructs that can be \textit{spread} directly from the parameter on which a \linear is built, to the \linear itself. These spreadable elements are of two kinds, functions, such as folds and map, and properties, such as the congruence of map or the satisfaction of a given predicate for at least one, or all, elements of the structure. We make use of the dependent type system of both \coq and \agda to model the latter. This paper ends with a discussion about
% We end by discussing 
various interesting topics that were raised throughout our development such as the issue of termination, the comparison of our tools and the proof effort required to extend \libName with additional elements.
\end{abstract}

\section{Introduction}\label{sec:intro}

Data structures are key in handling many programming challenges. From the easiest algorithms to more advanced features or conceptually challenging programs, choosing the right data structure is often mandatory in programming activities. Functional programming, and associated languages, usually provide constructs to model such structures, which can be summed up as \datatypes, where a type is defined with a list of constructors, each of which builds an element of the type  using a given number of inputs. While imperative and object-oriented languages do have \datatypes, they often manifest in a different manner, which will not be considered in this paper. These \datatypes are widely used in programming activities and can model various concepts, depending on the type system of the language in which they are defined. They can represent concrete data, in usual functional programming languages such as \caml and \haskell, and even properties in dependently typed languages such as \agda~\cite{agda} and \coq~\cite{coqart}, both of which have been used in this work.

As mentioned before, relying on relevant \datatypes to conduct programming activities is essential, even more so when said types are meant to model some high-level specification over concrete data. To help users in selecting the right \datatype, they are categorized and studied. Our work takes place in that area. The simplest \datatypes are the enumerations, where a type is built from a set of constant values, modelled by a set of unparametrized constructors.
More interesting \datatypes allow  constructors to have parameters. When one or more of these parameters are typed with the type that is being defined, the \datatype in question is recursive, a family of types which is all the more interesting. This family is particularly used and studied and is tied to the notion of induction. Among these recursive \datatypes, a distinction exists between regular \datatypes and non-regular, also called nested, \datatypes~\cite{thesis-bayley,bird-meertens-nested,ralf1998}.

A \datatype is \textit{regular} when its type parameter -- if any -- is always the same whenever it appears in the definition. In other words, if the type is polymorphic, then its type parameters are the same both in the signature of the type, as well as in the recursive constructors. Famous examples of regular \datatypes are  lists and trees, defined in \haskell as follows, where \texttt{a} is the type parameter:
\begin{Verbatim}
	data List a = Empty | Cons(a, List a)
	data Tree a = Tip a | Bin(Tree a, Tree a)
\end{Verbatim}

A \datatype is \textit{non-regular}, or \textit{nested}, when its type parameter changes between the type signature and at least one of its  instances in the constructors. Well-known nested \datatypes are the nests (sometimes called pow in the literature) and, most of all, the bushes, where the nesting (the way the type parameter changes) is done with the type definition itself, as shown below:
\begin{Verbatim}
	data Nest a = Zero a | Succ (Nest (a, a))
	data Bush a = BLeaf  | BNode (a, Bush (Bush a))
\end{Verbatim}

Nested \datatypes have been widely studied since 1998~\cite{bird-meertens-nested}, both theoretically using category theory, and practically in programming languages such as \haskell. Many approaches deal with nested \datatypes as an abstract notion, and try to theorize their use, regardless of their inner structure. Others are particularly interested in the folds that can be written on these structures~\cite{fold-bird-paterson, ralf1999a, fold-martin-gibbons-bayley, Fu2018, okasaki-1998}. Folding nested values is indeed mandatory because, due to the constrained nature of their type, this is the main way users have to interact with the value itself. Furthermore, these folds have to, and can, be as generic as possible, leading to the definition of powerful iterators, as well as induction principles. In his thesis~\cite{thesis-bayley}, Bayley is also interested in the genericity of other functions such as zip or membership. Our work shares this advocated approach of genericity around nested \datatypes.

In order to extend the number of notions that can be generic over nested \datatypes, we do not consider all their possible incarnations. Rather, we focus on a subset of all the possible nested \datatypes, which we call \textit{\linearsFull}. Moreover, we are interested in functions and properties that we can obtain for free from their definition, in the same spirit as \textit{Theorems for free}~\cite{th-for-free}, the deriving mechanism of \haskell~\cite{deriving-haskel}, or even the use of Finger Trees~\cite{hinze-paterson-2006}. Finger Tree is a general nested \datatype, parametrized by a monoid, the instantiation of which can lead to ordered sequences or interval trees. Indeed, our \linearsFull are characterized by a common structure with a changing type parameter, responsible for nesting the structure differently.

To that purpose, we have developed a core library, named \libName, available on the first author's github page\footnote{\gitURL}, in both \agda and \coq, making it accessible to a large number of users. This library provides the users with several nested \datatypes, defined as instances of \linears, as well as a core set of \textit{spreadable} components that are elements derivable  %have been derived 
from the type parameter to the nested \datatype itself. 
These spreadable elements are of two kinds, functions, such as folds and map, and properties, such as the congruence of map or the satisfaction of a given predicate for at least one, or all, elements of the structure. 
This paper presents the content of this library, with the following outline. 

Section~\ref{sec:lndts} presents the thought process behind the definition of our \linears, while also giving examples of \datatypes that can be built from them. Section~\ref{sec:computational} and~\ref{sec:logical} provide examples of spreadable constructs that can be derived for our \linears from the corresponding definitions of the underlying type parameter. Section~\ref{sec:computational} focuses on computational such aspects, while Section~\ref{sec:logical} focuses on logical ones. Section~\ref{sec:picture} shows a visual summary of these elements. Finally, Section~\ref{sec:discussion} proposes a discussion around some limitations and open questions that remain to be answered and that would benefit future works.

Throughout these sections, snippets of code are presented, which come from the \agda implementation of our work. These are type-checked pieces of code, ensuring their correctness, which is made possible through the use of \lagda, a tool to combine \agda code with \LaTeX~documents.

\section{Introducing Linked Nested DataTypes (\linears)}\label{sec:lndts}

\subsection{Introductory examples: \texttt{List$_0$}, \texttt{Nest$_0$} and \texttt{Bush$_0$}}\label{ssec:usuallnb}

The most common inductive \datatype is the type of lists which consists of an arbitrary number of elements linked one after the other. Written inductively using \agda, lists can be defined, as usual, using two constructors: the constant empty list \textcolor{Green}{[]} and the cons operator \textcolor{Green}{\_::\_} written in an infix manner using underscores. We use the subscript $_0$ to mean this definition is not derived from \linear.

\exampleList

Lists are parametrized by a given type \texttt{A} from a given level of universe \texttt{a}. An interesting feature of such a type -- that is usually not mentioned, although relevant in our case -- is that the recursive constructor takes as a parameter an element of type \texttt{List$_0$ A} where \texttt{List$_0$} is parametrized by the same type parameter as its definition, namely \texttt{A}. This makes it a regular \datatype rather than a nested one, where such a type parameter is assumed to vary throughout the recursion. As a first example of such a nested \datatype, let us consider the \texttt{Nest$_0$} \datatype: 

\exampleNest

In this case, the recursive constructor -- purposely named identically as the one for lists -- takes as a parameter an element of type \texttt{Nest$_0$ (A $\times$ A)} where \texttt{A $\times$ A} is a pair of elements of type \texttt{A}. This makes \texttt{Nest$_0$} a nested \datatype, where its type parameter evolves throughout the recursion using, in this case, the function \texttt{A → A $\times$ A}, which we call a \textit{type transformer}. As visible, both \texttt{List$_0$} and \texttt{Nest$_0$} are very similar in their structure, and their only difference is the type parameter on which the newly defined type is recursively called. As a last example, let us consider the \texttt{Bush$_0$} \datatype, nested with itself:

\exampleBush

In this case, not only does the type parameter changes in the recursive call, but it changes with a dependence to the type that is being defined. While picturing lists and nests is fairly simple, picturing a bush is challenging. Thankfully, while the parameter change depends on \texttt{Bush$_0$} itself, the form of the type is fairly similar to lists and nests, which calls out for a common denominator between the three -- and possibly more -- types, thus leading to a better picturing and understanding of bushes in the process. This is such a case where providing a relevant abstraction can significantly ease the study of its concrete counterparts, which is especially true for bushes, and motivates our approach.

\subsection{Definition of \linearsFull}\label{ssec:lndts}

The common denominator between the three aforementioned types is a structure we capture in our concept of \linears. \linears are parametrized by a \ttrans, that is an entity which, given a level of universe \texttt{a} and a type living in \texttt{Set a}, provides another type living in \texttt{Set a}. In \agda, our type transformers are called \texttt{TT}; they live in Set$\omega$ (a sort above all others), and are defined as follows:

\TTdef

This leads to the following \linear definition:

\lndt

It is interesting to note that this \datatype applied to a certain \ttrans is itself a \ttrans, that is, for any \texttt{F} of type \texttt{TT}, we have that \texttt{LNDT F} also is a  \texttt{TT}. Thus, LNDT is informally of type \texttt{TT $\rightarrow$ TT}\footnote{The signature of the related definition had to be tweaked a little to match Agda's distinction between type parameters and the sort of the datatype, a distinction marked by the colon.}. As a consequence \libName, in addition to defining spreadable properties over \linears, also introduces a certain number of \ttranss in the process, as shown in Figure~\ref{overview}.

\subsection{\texttt{List}, \texttt{Nest} and \texttt{Bush} as instances of \linear}\label{ssec:lndtlnb}

\paragraph{Tuples} Naturally, the three types considered as examples, \texttt{List$_0$}, \texttt{Nest$_0$} and \texttt{Bush$_0$} can be written as instances of \linear, once we provide the right \ttrans for each of them. While the \ttrans for \texttt{Bush$_0$} will be the type itself, we can notice that the \ttrans required for \texttt{List$_0$} and \texttt{Nest$_0$} can be abstracted in a notion that we call \texttt{Tuple}:

\tuples

A tuple indexed by \texttt{n} and parametrized by a type \texttt{A} is a collection of \texttt{n + 1} elements of type \texttt{A}. This is similar to the dependent types of vectors, where \texttt{Vec A n} stands for a list of \texttt{n} elements of type \texttt{A}. However, tuples are more convenient in our case because we never want them to be empty, which vectors can be, and they induce in our development some technical conveniences, on which we will not linger.

\paragraph{N-perfect trees} Having defined the tuples, the family of \linears based on them follows. They are parametrized by the size of the tuple on which they depend:

\nary

We can notice that \linears based on \texttt{Tuple} of a certain index \texttt{n} can actually be seen as ($n+1$)-perfect trees~\cite{ptree}. Any value for $n$ gives us a certain type of tree, with two examples being \texttt{List} and \texttt{Nest}, as defined in the following paragraph. They are perfect in the sense that all the nodes at a given depth either have no child or ($n+1$) children, which means that the overall number of nodes is $\sum\limits_{i=0}^k (n + 1)^i$, where $k$ is the depth of the tree.

\paragraph{Lists, nests and bushes} Thanks to the previous notions, we can finally jump to the definitions of \texttt{List}, \texttt{Nest} and \texttt{Bush}, seen as \linears with specific \ttranss:

\lndtdefs

This definition of \texttt{Bush} requires \agda to ignore termination checking -- although it is not shown here, it is required nevertheless --, which was not the case when defining \texttt{Bush$_0$}. This comes from the fact that positivity checking differs from termination checking. While the two definitions are equivalent, the first one relies on positivity checking because it is directly defined as a \datatype while the second one relies on termination checking since it is defined as an instance of \linear.

Such a termination checking cannot automatically succeed, which makes the use of \texttt{Bush} as well as functions over \texttt{Bush} and \texttt{Bush$_0$} unsafe. \coq rejects both definitions as unsafe and disallows their use, while \agda allows these definitions but considers them unsafe. This has the upside that working with bushes is possible in \agda, although the problem of termination checking to make their use safe remains and will be further discussed in Section~\ref{sec:discussion}.

\paragraph{Examples} Here is an example of \texttt{List}:

\examplelist

An example of \texttt{Nest}:

\examplenest

And an example of \texttt{Bush}:

\examplebush

\subsection{\texttt{Maybe} as instance of \linear}\label{ssec:maybe}

Another possible instance of \linear, is actually -- and surprisingly -- the usual \texttt{Maybe} type (\texttt{Option} in ML-like languages)
which is built from the \texttt{Null} \ttrans, corresponding to the logical negation.

\lndtnull

The idea comes from the fact that such a \texttt{Null} \ttrans always builds empty types, thus only allowing the structure to be empty (using \texttt{[]}), or to contain a single element (using \texttt{\_∷\_}), since any further nesting will not be inhabited.

In addition, by using the \texttt{pattern} keyword provided by \agda, which allows the developer to define aliases usable in pattern-matching situations, we can provide an alternate definition of the \texttt{Maybe} type as an instance of \linear, by mapping \texttt{nothing} to \texttt{[]} and (\texttt{just x}) to (\texttt{x ∷ []}). By doing so, we ensure that this definition and the usual one are syntactically equivalent, in addition to behaving the same way.

\maybe

This type has the right semantics, which means that any element of type \texttt{Maybe} is either of the form \texttt{nothing} or \texttt{just x}, as expected. This has been proven in \libName although the proof is not presented here. While defining \texttt{Maybe} in this alternate manner is not necessarily relevant \textit{per se}, it is interesting to see another way of building such a type, as well as to notice that \agda, thanks to the \texttt{pattern} mechanism, allows the developer to use this type exactly as one would use the usual \texttt{Maybe} type from the current \agda standard library. Moreover, discovering hidden patterns between types is often enlightening, and always exciting.

\subsection{Multi-layered \linears}\label{ssec:morelndt}

As noticed in Section~\ref{ssec:lndts}, for any \texttt{F} of type \texttt{TT}, \texttt{LNDT F} has type \texttt{TT} too, which means that we can keep building new interesting \linears by chaining multiple calls to \texttt{LNDT}. While we say "interesting", it is fair to say that not all such attempts indeed bear that characteristic. However, some do, and below is an example of a second degree \linear that can be built, and from which multiple features will be retrieved for free, thanks to the spreadable properties depicted later on. We call this type \texttt{SquaredList}:

\lndtlist

Here is an example of an inhabitant of \texttt{SquaredList}, starting with a natural number, then a list of natural numbers, then a list of lists of natural numbers, and so on:

\lndtlistexample

\subsection{Overview of our \linears}\label{ssec:overviewlndt}

\begin{figure}[h!]
\centering
\includegraphics[width=\textwidth]{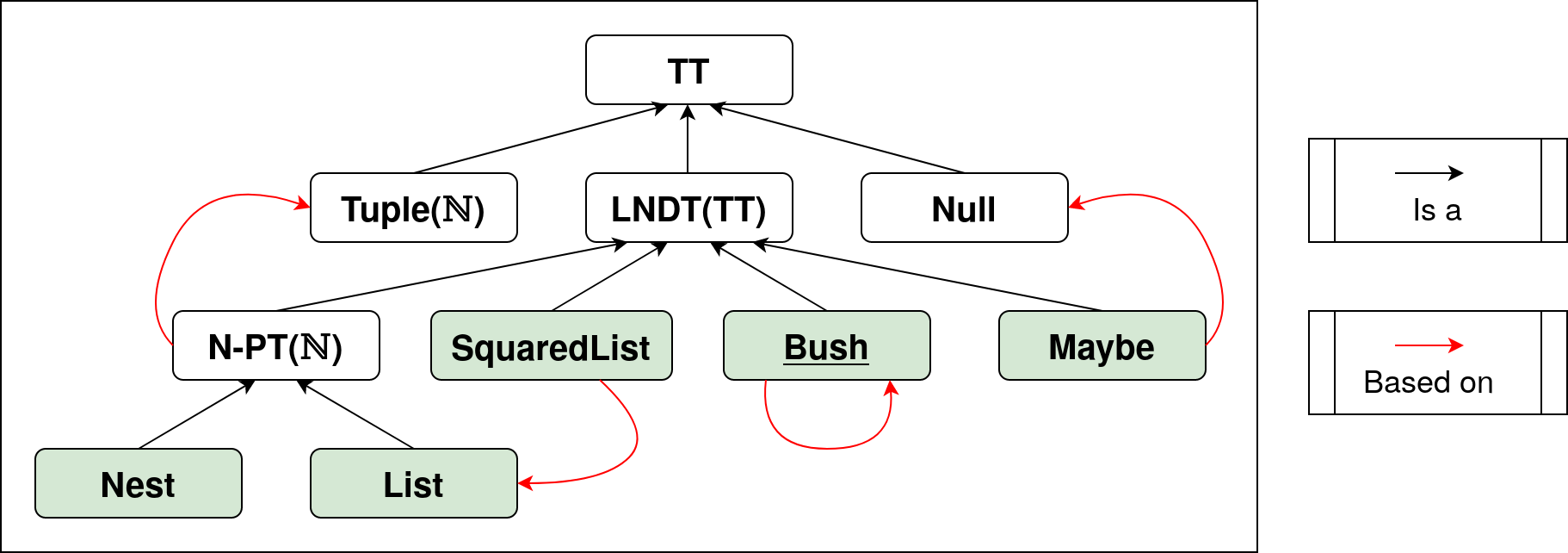}
\caption{Type transformers in \textsc{LibNDT}}
\label{overview}
\end{figure}

Figure~\ref{overview} shows an overview of the \ttranss that were defined in \textsc{LibNDT}, and how they relate with one another. The light green boxes represent concrete types while the others are abstract types which were used to build them. For instance, the box labelled "Maybe" and the arrows that depart from it, depict that the type \texttt{Maybe} is the result of the instantiation of \texttt{LNDT} with \texttt{Null} as a parameter. In the case of the \texttt{Bush} type, the label of the related box is underlined to indicate that the definition is not safe.

\section{Computational common behaviours}\label{sec:computational}

However tempting regrouping these types under a common denominator is, the relevance of this process depends on the possibility to express behaviours at the \linear level which could then be instantiated to any of its instances. The remaining of this paper stands as a list of arguments in favour of this relevance, that is, a list of behaviours that can indeed be expressed for \linears. Such behaviours will be regrouped into two classes: \textit{computational} ones, that is concrete functions relying on the content of \linears, and \textit{logical} ones, that is properties either directly bound to \linears or functions that work on them. In other words, the first class contains functions that process elements from \linears while the second class contains everything else. This section depicts our results regarding the first class of common behaviours.

\subsection{Mapping \linears}\label{ssec:maps}

When considering functions that work on collections of elements, a few examples come to mind, the first of which is the \texttt{map} primitive. The reasoning behind \texttt{map} is well-known and natural: when possible, using some function \texttt{f}, to transform elements of type \texttt{A} to elements of type \texttt{B}, it should be possible to transform a collection of elements of type \texttt{A} to a collection of elements of type \texttt{B} through a procedure, parametrized by \texttt{f}, called \texttt{map}. We begin by studying this assumption for \linears.

\paragraph{Building map functions for \linears} As this is our first example of common behaviour, let us consider, in details, the steps in our reasoning. Since lists are the simplest example of \linear, we can start by considering the usual \texttt{map} function written on lists. Below is the common definition of maps for lists (subscripted by $_0$), which bears no understanding difficulties. 

\listmap

Writing a similar function for nests is more challenging because the tail of the nest is not of the same type of the nest itself, which means the same function \texttt{f} cannot be passed as it is in the recursive call.

\nestmap

However different, these two have a lot in common, which can be captured using \linears, as long as the way the function \texttt{f} must be transformed throughout the recursion is provided. By taking as parameter this transformation of function, and calling it \texttt{T}, we propose the following implementation for \linears.

\lndtmapz

While the signature of this function is not straightforward, it is possible to make it clearer by noticing a certain regularity in its core, which can be made visible as follows:

\mappable

This definition gives an abstract signature to \ttranss \texttt{F} for which it is possible, given a function from \texttt{A} to \texttt{B}, to build a function from \texttt{F A} to \texttt{F B} that is, a map function. Using this new notation, the map function over \linears has a far better looking -- and much more explicit -- signature.

\lndtmap

The latter definition gives us an important insight as to what \texttt{lndt-map} stands for: it is a procedure that ensures it is possible to map on \linears provided it is possible to map over the \ttrans on which they are build. This is the first property that justifies the spreadable aspect of this library: we study which elements can be transported from \texttt{F} to \texttt{LNDT F}, and \texttt{Map} as defined earlier is one of them.

\paragraph{Instantiating map functions} In order to deduce map functions for \linears, all is needed is to define a map function for the \ttrans on which they are based. The first function, which can be spread to the corresponding \linear, is called the \textit{seed} for the associated behaviour. In this case, the behaviour is the ability to map over a given type. Here is the map seed for tuples:

\tuplesmap

The definitions of maps for \linears are hence straightforward:

\lndtmaps

We can notice that the map function on bushes is recursively generated using itself, solely relying on \texttt{lndt-map} as a way of computing a result. In other words, the seeds for bushes will never have to be defined and any behaviour on bushes is always solely generated with the associated spreadable property.

\paragraph{Examples of usage of map functions} As a first example, here is a list of natural numbers on which the successor function is mapped:

\firstexamplelist

As a second example, we define a bush of natural numbers on which we map the function of multiplication by two:

\firstexamplebush

In both cases, \texttt{refl} is a correctly typed term regarding the signature of the function, which means both sides of the equality are indeed the same. These examples are by no means a proof of correctness of the \texttt{map} functions, but rather a convincing argument in favour of our approach and definitions.

\paragraph{Back to squared lists} Squared lists are second degree \linears, in the sense that they are built by  two successive  nestings. However, regardless of the number of successive nestings, we can provide a map function as long as the seed -- the original \ttrans -- provides a map function itself. The map function  for squared lists is obtained as follows:

\lndtlistmap

As an example of usage of this newly created map function, we can apply a multiplication by two on all elements of the example %squared lists 
defined in Section~\ref{ssec:morelndt}.

\lndtlistmapexample

As shown in this example, nesting several times over a given \ttrans does not alter our ability to provide free functions from the seed of the chain. Throughout this paper, more examples of spreadable behaviours will be given, all of which can be spread several times similarly. As a consequence, we will not explicitly go back to squared lists in the rest of the paper.

\subsection{Folding \linears}\label{ssec:folds}

Another common behaviour which can be defined over \linears directly are the fold functions, right or left, which walk through the structure while combining their content using a given operator and a given seed. While we described the whole thought process behind the map function, we will, from now on, only give the property that is spreadable and the associated definitions. Both folds are spreadable and can thus be transported from \texttt{F} to \texttt{LNDT F}. Both folds share the same abstract type:

\fold

Then, assuming we can fold over the type parameter, we can propagate this fold to \linear in two ways, left or right. Below is the left propagation, while the right one is also present in the library:

\folds

As an example of left fold on an inner \ttrans, we define the left fold on tuples:

\foldtuples

This leads to the definition of folds for our \linears. Here are the left folds for bushes and nests, with the respective seeds the fold on tuples, and itself.

\foldsinstances

Here is an example of left folding a nest of strings:

\foldsexampleone

And an example of left folding a bush of strings:

\foldsexampletwo

\subsection{Summary}\label{ssec:computationalsummary}

This section exhibited three spreadable elements, that can be built \textit{freely} for \texttt{LNDT F} when they exist on \texttt{F}, these are the two folds, left and right, and the map function. These elements can be regrouped inside a record which contains all spreadable properties, which we call \texttt{SpreadAble}, and which will be enriched with logical properties in the next section. From these elements, it is possible to build additionnal functions directly, such as \texttt{size} (the number of elements contained in a specific structure) and \texttt{flatten}, returning a list of these elements. This has been done in the library but is not shown here.

\section{Logical common behaviours}\label{sec:logical}

Until now, we considered functions on \linears which provide a concrete value from such types, without any type dependence to values of any kind. In other words, most of what has been shown earlier could have equally been developed in a classical functional programming language with polymorphic types. Yet, we work with dependent types, which enlarges the boundaries of what can be expressed around our \linears. This section shows examples of what we call logical properties, which means any definition about \linears whose type is dependent. They include primitive predicates over \linears, predicates around the computational aspects of \libName alongside decidability properties.

\subsection{Primitive predicates for \linears}

\subsubsection{Predicate transformers}

Our first idea is to express the satisfaction of a predicate by all elements, or at least one element inside our \linears. In other words, we want to define, given a predicate \texttt{P}, the usual predicates \texttt{All P} and \texttt{Any P} over any \linear.

The first step in that direction is to define the type of predicate transformer, transforming a predicate over a certain type \texttt{A} to a predicate over \texttt{F A}, where \texttt{F} is a certain \ttrans. The definition is as follows, where \texttt{Pred} is the \agda type for unary predicates (\texttt{Pred A b = A → Set b}):

\transpred

This specification now defined, we need to find inhabitants for them on our \linears, the semantics of which should be respectively \texttt{Any} and \texttt{All}. Both have been defined and can be found in the library, however, we only show \texttt{Any} in details here. Since a \linear is defined using a certain \ttrans as a parameter, this \ttrans needs to be associated with a predicate transformer itself, so that our \linear can provide its extended version. In other words, we look from an inhabitant of \texttt{TransPred (LNDT F)} provided we have an inhabitant of \texttt{TransPred F}. This is built as an inductive \datatype, similarly as the usual definition of \texttt{Any} over lists, for instance.

\transany

There are two cases, either the first element of the structure satisfies \texttt{P}, or one of the elements of the tail of the structure satisfies \texttt{P} nested with \texttt{T}, which is the predicate transformer associated with the underlying \ttrans. In order to give examples, we define the predicate transformer \texttt{Any} over tuples, so that it can be propagated to nests and lists.

\tupleany

This definition uses \texttt{⊎} which represents in \agda the logical union and consists of two cases: either the tuple contains a single element, in which case \texttt{P} itself is returned, or it contains more that one, in which case either the first element satisfies \texttt{P}, or one of the others satisfies it recursively. From this, we define \texttt{Any} for nests and bushes, relying on \texttt{Any} on tuples and itself respectively.

\anyinstances

Here is an example of \texttt{Any} over bushes, with the proof that the number $10$ is a member of a bush. It can be found somewhere in the third bush, following the chain of \texttt{here} and \texttt{there} from \texttt{lndt-any}.

\anybushexample

\subsubsection{Decidability transformers}

Since we are able to propagate a predicate transformer, we would like to tackle the propagation of the decidability of predicates. To do so, we express what it means for a predicate transformer to preserve decidability.

\transdec

From this definition, we can prove that our predicate transformers \texttt{Any} and \texttt{All} over \linears do preserve decidability on the condition that the underlying predicate transformer does so. Proofs are omitted, although here is the signature relative to \texttt{lndt-any}:

\lndtdecany

More concretely, this means that in our library, the decidability of a predicate \texttt{P} can be propagated to the decidability of \texttt{Any} and \texttt{All} for all our \linears. In other words, we are able to decide if at least one, or all elements of a \linear do satisfy \texttt{P}, even for bushes.

\subsection{Predicates over computational aspects}

A second logical family over \linears revolves around the satisfaction of predicates for the functions we have defined. This is a wide area that consists in giving specification for our functions and proving that they %said function 
do satisfy their specification. Although our functions are low-level and not composite, which makes the specification process all the harder, we have several such examples in our library, despite only one being presented here, the congruence of a mapping. In other words, we present the proof that mapping with a function \texttt{f} is the same as mapping with a function \texttt{g} on any \linears provided these function coincide on every input (aka are extensionally equal). We start by defining this abstract property.

\mapcongruence

We provide the proof that our \linears preserve the congruence of a mapping. It means that, if 
%a certain mapping 
\texttt{map} is congruent over a \ttrans \texttt{F} then \texttt{lndt-map map} is congruent over \texttt{LNDT F}. The proof is done inductively, and is shown here as an example of such a proof in \agda, although hardly understandable for non \agda users, admittedly.

\mapcongproof

\subsection{Summary}

This section provided examples of logical properties expressed around our \linears. We showed that we can build predicates over \linears that stand for the membership modulo a predicate, named \texttt{Any}, and the satisfaction of a predicate for all members of a \linear, named \texttt{All}. We showed that these were decidable provided the unary predicate over their elements is decidable. We also showed an example of specification for one of our computational behaviour:  mapping with a given function, which stays congruent. All these elements were built with the same underlying idea  used throughout this paper: the propagation of properties from \texttt{F} to \texttt{LNDT F} where \texttt{F} is a \ttrans. Next section summarizes all our definitions in a single picture.

\section{Picturing the spreadable elements of \libName}\label{sec:picture}

	\begin{figure}[h]
		\centering
		\includegraphics[width=\textwidth]{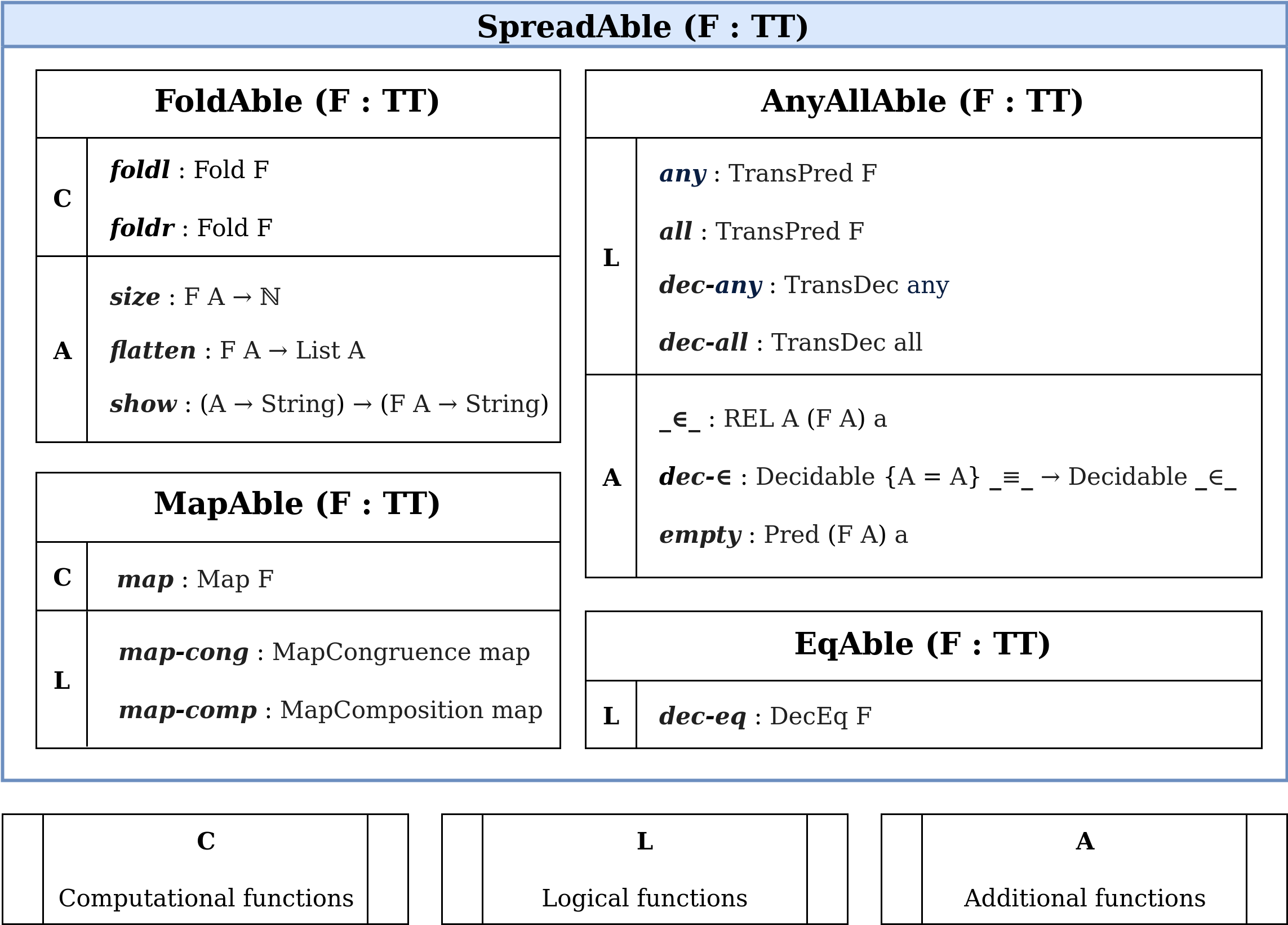}
		\caption{Overview of the library \libName}
	\end{figure}

Our spreadable properties, most of which have been mentioned throughout this paper, are regrouped into four categories, each corresponding to a concrete data structure in \libName, modelled as a record:
\begin{enumerate}
\item \texttt{FoldAble} contains the two folds, left and right, with three additionnal functions, \texttt{size}, \texttt{flatten} and \texttt{show} which, respectively, counts the number of elements in our structure, flattens the structure into a list, and produces a string representation of the structure.
\item \texttt{MapAble} contains a map function, with two related properties, the congruence and the composition associated with it
\item \texttt{EqAble} contains the proof of decidability of equality between \linears.
\item \texttt{AnyAllAble} contains two predicate extensions over \linears, the ``at least one element'', and the ``all elements'' satisfying a given underlying predicate, alongside their decidability and additional derived constructs: the membership in a \linear, its decidability, and the emptiness of a \linear.
\end{enumerate}
All these elements are regrouped into a \texttt{SpreadAble} record which contains all spreadable elements of the library, and thus is characterized by \texttt{SpreadAble F → SpreadAble (LNDT F)}.

\section{Discussion}\label{sec:discussion}

This last section brings together a set of observations and open questions that remain to be discussed and that we would like to underline. Most of these elements have been mentioned throughout the paper and need additional explanation. They are displayed in no particular order.

\paragraph{\coq and \agda} Our development was presented in \agda, but is also available in \coq, as mentioned earlier. Having these two implementations of our ideas was fruitful in terms of providing users with our contribution, as well as in noticing differences between the tools. The main one lies in the definition of \texttt{Bush} which is accepted by \agda and refused by \coq, regardless of the use of \linear to define it (\texttt{Bush}) or not (\texttt{Bush₀}). This means that \coq is, in that regard, more restrictive than \agda in the sense that it refuses any definition from which subsequent functions would be troublesome in terms of termination proofs. \agda accepts these definitions -- \texttt{Bush₀} directly, \texttt{Bush} when termination checking is disabled -- although subsequent definitions do indeed induce termination proofs issues in both cases. The good thing about being able to define \texttt{Bush} is that we can work with them by forcing the termination checker to trust us, although this leads to an unsafe development. Working in this manner is of course discouraged when dealing with safety issues, but it can prove to be interesting to handle this hard-to-grasp type. In our work, all functions that work on \texttt{Bush} do so in such a way that they are recursively defined using themselves, with a base case hard to picture, although they produce relevant results. This would be interesting to see how and if \coq could be extended to accept similar definitions, to what extend, and at which cost.

\paragraph{The issue of termination} As mentioned earlier, \texttt{Bush} is troublesome in terms of termination, which is unsurprising considering the nature of this datatype. In that regard, languages in which bushes can be used do not usually embed a termination checker and instead rely on trusting the developers ; and when they do like in \agda, it needs to be turned off for the time being. An open question is to what extend such a termination could be proven. Considering the definition of \texttt{Bush} provided in Section~\ref{ssec:lndtlnb}, \texttt{Bush = LNDT Bush}, it is highly unlikely that any termination checker would ever accept this definition, and that any of the usual termination techniques (exhibiting a well-founded order, using sized types, etc.) would allow to extend it in a way that they do. To tackle this issue, it is possible to give an alternate, more verbose definition of \texttt{Bush} using an indexation of the number of times \texttt{Bush} is encapsulated. This definition is accepted by \agda but also by \coq, as shown by the following snippet:
\begin{lstlisting}[language=Coq]
Inductive BushN : nat -> Type -> Type :=
 | Base (a : Set) : a -> BushN O a
 | NilBN (a : Set) (n : nat) : BushN (S n) a
 | ConsBN (a : Set) (n : nat) : 
          BushN n a -> BushN (S (S n)) a -> BushN (S n) a.
\end{lstlisting}
This was tested in our library, but it remains unclear to which extend this solves the termination issue while providing the same expressiveness. Furthermore, such a definition does not comply with our overall pattern captured by the notion of \linear. Overall, the termination issue over \texttt{Bush} remains open.

\paragraph{Automated term generation} \agda comes with an automated term generator, named \agsy, which, when called, attempts to build a term in a certain context with respect to a certain goal. \agsy works fine with \libName, except for \texttt{Bush} where termination issues appear in the process of term generation. More precisely, while functions on \texttt{Bush}, although we are not able to prove it, do terminate, automated term generation over \texttt{Bush} does not.  As far as we could observe, this behaviour might come from \agsy's heuristic to explore possible terms with the maximum number of elements, rather than the opposite. This would however be interesting to investigate deeper why this happens, and if this would be relevant to implement different heuristics inside \agsy and similar automated theorem provers, to fix this issue.

\paragraph{Extending \libName with more spreadable elements} \libName provides a representation of several \datatypes using our notion of \linears, all of them consisting in a collection of elements structured in a head-tail manner. Each of the original types that we model as instances of \linear comes in the literature with a set of candidate functions for abstraction over \linear. Most of the usual functions coming from \texttt{List}, for instance, have been considered as possible functions to define for \linears as a whole. Currently, \libName consists of several \linears, on which 16 functions (either computational, logical or additionnal) can be derived automatically from the underlying \ttrans to the corresponding \linear. Possibly, many more functions could be added to this set, although a lot of them have been considered and ultimately proved impossible to abstract. An example of such a case is the idea of defining \texttt{map} using \texttt{fold} which seems both appealing, because it would reduce the minimal bricks of our \libName, and promising, since \texttt{map} can indeed be written using \texttt{fold} when considering \texttt{List}. However, in practice, this is impossible due to type requirements. Indeed, the intuitive way of building \texttt{map} using \texttt{fold}, which works for lists, is the following:
\begin{Verbatim}
	map f = foldr (λ a l → f a ∷ l) []
\end{Verbatim}
This does not transpose into any \linear because, \texttt{l} is of type \texttt{LNDT (F B)} although \texttt{f a ∷\_} expects an element of type \texttt{LNDT (F (F B))}, where \texttt{f} is of type \texttt{A → B}. Since \texttt{F} is the identity for \texttt{List}, it works, but it is a special case and can only be generalized for any \texttt{F} such as \texttt{F ∘ F = F}. The common ancestor of both \texttt{map} and \texttt{fold}, when considering \linears, is the induction principle derived from the definition, and one cannot be written using the other. As a side note, this induction principle is successfully generated by \coq, and can be written in \agda.

Another example of failed attempt at abstraction is the \texttt{zip} function, which cannot be extended to \linears because the structures need to be similar to be zipped together, and for instance any two bushes have little chance to be similar enough. We are confident, however, that more functions could be abstracted, especially in the logical area. For instance, we were interested in specifying contracts on our computation functions and were wondering if we could for instance prove, using our notion of membership and our notion of mapping that, given a \linear \texttt{l} and an element \texttt{x} such that \texttt{x} is a member of \texttt{l}, the fact that \texttt{f x} is a member of \texttt{map f x}. This should hold, however, it proved to be hard both to express and to prove. We are confident that this can be proven in our library; however, this requires a strict discipline on how to reliably find invariants in recursive function over \linears, when the signature of the function contains elements that are not concerned by this recursivity. This remains an open question and will most likely be the subject of future work.

\paragraph{Extending \libName with additional \datatypes} Since, by nature, the work presented in this article only applies to a specific subset of nested \datatypes, there are some other nested \datatypes which are not handled in our work. An example are the well-known Finger Trees~\cite{hinze-paterson-2006} which require an additional constructor to be defined -- although some tricks using \texttt{Null} and \texttt{Maybe} could be considered. This observation leads to interesting questions as to how our work could be extended to nested \datatypes that do not directly satisfy the structure we propose in \linear. Such questions have been tackled over regular \datatypes in different ways, which would be interesting to consider when relying on our work to use nested \datatypes. These possibilities are as follows:

\begin{enumerate}
\item A first possibility is to rely on even higher abstractions, which means studying nested \datatypes as an abstract notion rather than a concrete family of inductive types. This is possible and has been done in other works such as~\cite{thesis-bayley}. However, this suffers from the usual drawback of high level abstraction: they are very far from concrete preoccupations and thus can hardly be used to obtain free code for their instances. In our case, the structure depicted in \linear is essential to any of the concrete elements we have, which enforces a strong confidence in the level of abstraction we chose.
\item A second possibility is to resort to meta-programming, which exists in \coq with frameworks such as ``Coq à la carte''~\cite{meta-carte-2} or more recently ``MetaCoq''~\cite{Sozeau20} and which is currently under development in \agda. Such a meta-programming would allow us to define new \datatypes at runtime, but it is not yet clear to us if this would allow us to reuse some of our result with little effort, if any.
\item A third promising possibility would be to use ornaments~\cite{ornaments-in-practice, dagand-mcbride-2014}. Ornaments are a way to build a hierarchy between \datatypes from which functions can be derived with a certain degree of automation. Using ornaments would possibly allow \libName to handle many more nested \datatypes.
\end{enumerate}

\section{Conclusion}\label{sec:conclusion}

\paragraph{Assessments} In this paper, we have defined a restricted class of nested \datatypes with a similar structure, the \linears. All different instances of \linear are built using a different parameter called a \ttrans. We studied these types over two axes: first, we studied a family of types that can successfully be seen as a \linear, and second, we built a set of functions, whether computation or logical, which can be derived automatically from the underlying \ttrans to the corresponding \linear. Throughout this investigation, we developed a library regrouping these elements, called \libName. This library was developed both in \coq and in \agda. The main difference between the two implementations is that \agda allowed us to define the type \texttt{Bush}, on which all our functions can be applied. \libName is a small library, which contains the most fundamental notions to working with \linears, and which will be extended with new elements in the future.

\paragraph{Limitations} In that regard, our work has several limitations, most of which have already been discussed in Section~\ref{sec:discussion}, and can be summarized as follows. Our work is focused on a specific subset of nested \datatypes, which is thus less generic than other higher level approaches, although it provides a bigger operational part. Our library could be used to model even more \datatypes, and could be extended with additional spreadable properties. Moreover, our library does allow the user access to the \texttt{Bush} type when using \agda, but this use is unsafe by nature, since termination issues are yet to be tackled. Finally, our approach provides a generic framework over \linears, although we do not provide a way to export the notions it contains to other kinds of nested \datatypes.

\paragraph{Perspectives} These limitations bring perspectives to our work. We would like to extend \libName to cover a wider ground in terms of possible instances of \linear, as well as in terms of possible spreadable properties. We would also like to extend \libName to handle more kinds of \datatypes. This could be done using meta-programming or ornaments. We would also like to investigate some termination issues, based on \texttt{Bush} but not limited to. Indeed, our work contributes to raising the question of proofs of termination regarding recursive function when the defined function is passed as parameter inside its body to another function. Finally, we would like to investigate to which extend proof assistants should accept the type \texttt{Bush} and similar types, and the cost of such an acceptance.

\newpage

\bibliography{msfp2022}
\end{document}